\documentclass[fleqn,usenatbib]{rasti}
\usepackage{newtxtext,newtxmath}
\usepackage[T1]{fontenc}

\DeclareRobustCommand{\VAN}[3]{#2}
\let\VANthebibliography\thebibliography
\def\thebibliography{\DeclareRobustCommand{\VAN}[3]{##3}\VANthebibliography}

\usepackage{graphicx}	%
\usepackage{amsmath}	%

\title[Transformer models for astrophysical time series]{Transformer models for astrophysical time series and the GRB prompt-afterglow relation}

\author[O. M. Boersma et al.]{
Oliver M. Boersma,$^{1}$\thanks{E-mail: o.m.boersma@uva.nl}
Eliot H. Ayache,$^{2}$
Joeri van Leeuwen$^{3}$
\\
$^{1}$Anton Pannekoek Institute for Astronomy, University of Amsterdam, Science Park 904, 1098 XH Amsterdam, The Netherlands \\
$^{2}$The Oskar Klein Centre, Department of Astronomy, Stockholm University, AlbaNova, SE-106 91 Stockholm, Sweden\\
$^{3}$ASTRON, Netherlands Institute for Radio Astronomy, Oude Hoogeveensedijk 4, 7991 PD Dwingeloo, The Netherlands
}

\date{Accepted 13-June-2024. Received 26-Apr-2024; in original form 11-Jan-2024}

\pubyear{2024}

\begin{document}
\label{firstpage}
\pagerange{\pageref{firstpage}--\pageref{lastpage}}
\maketitle

\begin{abstract}
Transformer models have recently become very successful in the natural language domain.  Their value as
sequence-to-sequence translators there, also makes them a highly interesting technique for learning relationships
between astrophysical time series.  Our aim is investigate how well such a transformer neural network can establish
causal temporal relations between different channels of a single-source signal.  We thus apply a transformer model to
the two phases of Gamma-Ray Bursts (GRBs), {reconstructing one phase from the other}.  GRBs are unique instances where a
single process and event produces two distinct time variable phenomena: the prompt emission and the afterglow.  We here
investigate if a transformer model can predict the afterglow flux from the prompt emission.  If successful, such a
predictive scheme might then be distilled to the most important underlying physics drivers in the future.  We combine
the transformer model with a novel dense neural network setup to directly estimate the starting value of the prediction.
We find that the transformer model can, in some instances, successfully predict different phases of canonical
afterglows, including the plateau phase. Hence it is a useful and promising new astrophysical analysis technique.  For
the GRB test case, the method marginally exceeds the baseline model overall, but still achieves accurate recovery of the
prompt-afterglow fluence-fluence correlation in reconstructed light curves. Despite this progress, we conclude that
consistent improvement over the baseline model is not yet achieved for the GRB case.  We discuss the future improvements
in data and modeling that are required to identify new physical-relation parameters or new insights into the single
process driving both GRB phases.
\end{abstract}


\section{Introduction}
\label{Transformer_Intro}
Transformer neural networks~\citep{NIPS2017_3f5ee243} have found great success in recent years in the machine learning domain. This type of neural network originates from natural language processing (NLP) needs. It excels when used as a building block of large language models like GPT-3~\citep{GPT3}, producing state-of-the-art performance. Moreover, transformer models have been applied with favourable results to tasks such as computer vision (Vision Transformer,~\citealt{2020arXiv201011929D}), time-series forecasting~(e.g.,~\citealt{2022arXiv220207125W}), and protein structure determination~(e.g.,~\citealt{rao2021transformer}). Transformer models have  been utilized for problems in astronomy as well
(e.g.,~\citealt{2021arXiv210506178A,Jia_2023,10.1093/mnras/stad1173}), but one could argue that their use has not been as
widespread as other deep learning models such as standard fully connected dense neural networks or convolutional neural
networks (e.g.,~\citealt{2018AJ....156..256C,10.1093/mnras/stz1288,2022ExA....53....1S}).

The attention mechanism at the heart of transformer models is key to their success, giving context both to different tokens
solely in the input or output sequence (self-attention) as well as between tokens in the input and output
(cross-attention). In contrast to recurrent neural networks such as long short-term memory ~\citep[LSTM;
][]{2014arXiv1409.3215S}, transformer models process the entire input at once instead of in a sequential fashion. {Convolutional neural networks (e.g.,~\citealt{Gu2015RecentAI}) also process data in parallel within each convolution layer, but their potential to capture long-range dependencies is limited by the kernel size. This necessitates the use of a number of convolutional layers which decreases efficiency.} The {more direct}
parallelisation {of transformer models captures long-range dependencies efficiently and} has the ability to greatly reduce training time, allowing for much larger datasets to be used, although
this still comes with significant training cost. Even so, in NLP-related tasks like machine translation, where big
datasets are rather easily obtained, transformer models have become the de facto standard model, trained in parallel on large numbers of graphical processing or tensor processing units.

We here apply transformer models to gamma-ray bursts (GRBs), {treating the two phases of GRBs as a translation problem, attempting to derive the latter phase from the earlier phase}. These bright flashes of intense gamma radiation are produced in relativistic blast waves through interactions and shock waves within the jet. When the matter of these
expanding explosions hits the surrounding medium, an afterglow is produced through synchrotron radiation spanning the
electromagnetic spectrum. GRBs come in two types, long and short \citep{1993ApJ...413L.101K}, each with its own
leading model for the origin. Long-duration bursts are {associated with} the collapse of massive stars to black holes, or magnetars~(e.g.,~\citealt{BERNARDINI201564}), after
a supernova \citep{1998Natur.395..670G}, while short-duration bursts are {linked to} the merger of compact objects like neutron stars \citep{2006Natur.444.1044G}. {It is often posited that black hole hyperaccretion through neutrino-dominated accretion flows~(e.g.,~\citealt{1993ApJ...405..273W, popham1999hyperaccreting, liu2017neutrino}) or neutron star spin-down~(e.g.,~\citealt{1992Natur.357..472U, Zhang_2001}) are the central engines of GRBs, although a wider range of theories exists~(see e.g.,~\citealt{meszaros2002theories, zhang2004gamma, berger2014short}).}

A central topic in GRB astronomy is how the prompt emission of gamma radiation and the afterglow emission are related
(Sect.~\ref{sec:Transformer_GRBphysics}). The models for the prompt and afterglow emission naturally imply a causal connection between the two and this has readily been observed.~\citet{Gehrels_2008}, for example, find that strong bursts generally have brighter afterglows and that short bursts are weaker in both prompt and afterglow emission.~\citet{Liang_2005} find a similar but different parameter correlation between the isotropic energy of the prompt emission, the intrinsic prompt emission peak energy, and the jet break time of the optical afterglow. 

In this work, instead of studying inferred parameters
of the GRB prompt and afterglow emission
to find a connection and express it in a number of summary metrics,
we use machine learning
methods to identify the relationship between the two astrophysical time-series. The transformer model, in particular, seems well suited to this task as it can be thought of as a sequence-to-sequence
translation problem similar to that of, for example, language translation. {We hypothesise that their strength in maintaining a consistent translation over entire paragraphs of text~\citep{NIPS2017_3f5ee243} is well suited to capturing long-range relations in GRB time series too}.
We train the transformer model to directly translate light curves of the prompt emission to the corresponding light curves of the afterglow emission. We make use of the only moderately big dataset\footnote{Between acceptance and publication, 
the code for data acquisition and analysis
will be made available at: \url{http://doi.org/10.5281/zenodo.10932872}} 
available for combined GRB prompt and afterglow measurements, consisting of GRBs and their afterglows detected by the Swift satellite over the past 20 years~\citep{2004ApJ...611.1005G}.

The main motivation undertaking this work is to look for evidence that  links the prompt and afterglow emission
--  without a preconceived notion of which physics is involved. If any predictive scheme is found, subsequent analysis
could then distil this relation into more specific physical insights, {leading to new insights into their physical origin. If the transformer model finds, for example, that a strong link exists between specific parts  of the prompt and afterglow light curves separated by time $t$, future work could investigate 
which process delays the energy injection by this time $t$, 
and what the implied typical scale sizes and jet velocities mean
for the origin of GRBs 
(see Sect.~\ref{sec:Transformer_GRBphysics} for further possible research directions and prospects).
} The latter analysis is, however, not yet part
of this work. Instead, we here first investigate solely the ability of the transformer model to establish causal
temporal relations between different channels of a single-source signal. The prompt and afterglow emission channels of
GRBs lend themselves well to such methodology, as we know a priori some relation must exist; but to our knowledge
this relation has thus far always been approached from a physics-first point of view. Our aim here is to tackle this relation from an empirical machine-learning perspective, and provide a new way of strengthening the prompt-afterglow connection. Additionally, if we do not find a significant link between prompt and afterglow behavior through this method, a strong connection between the two emission types could be disproven.

{This paper is organized as follows.} Section~\ref{sec:Transformer_GRBphysics} provides a short overview of the theory behind the physics connecting the prompt and afterglow phase of GRB emission. In Sect.~\ref{sec:Transformer_Data}, we go into the specifics of the chosen Swift data set. We give an overview of our tailor-made transformer model in Sect.~\ref{sec:Transformer_NN} and present the results in Sect.~\ref{sec:Transformer_results}. We discuss our findings in Sect.~\ref{sec:Transformer_disc} and conclude in Sect.~\ref{sec:Transformer_conc}.

\section{Science Case: Connecting the two types of GRB emission}
\label{sec:Transformer_GRBphysics}
There is no consensus yet on the exact physical explanation for the parameter correlations between the prompt and
afterglow emission of GRBs (see~\citealt{DAINOTTI201723} for a recent review). Although the total energy reservoir from
the central engine is shared between the prompt and afterglow phases, predicting the exact partitioning is not straightforward due
to the complex nature of the jet structure, the medium density, the shock microphysics, and the emission
processes. %
Still, as the prompt and afterglow emission originate in a single source,
we expect there to be a connection (a correlation or anti-correlation) between their light curves.

The rapid variability in the prompt light curves indicates an internal origin close to the compact
object, i.e., a hot fireball expanding adiabatically \citep{1992MNRAS.258P..41R};
the slower decay phases of the afterglow, on the other hand, suggest
interactions with an external medium.
The prompt emission is produced by the dissipation of
kinetic energy via internal shocks within the jet \citep[see, e.g.,][]{1997ApJ...490...92K}.
In early afterglow observations, often a steep decay phase is present which is interpreted as the late signature of the
high-latitude (off-axis) prompt emission.
Following this steep decay, many GRB afterglows exhibit a plateau with
very shallow decay \citep[see, e.g.,][]{2005ApJ...625L..23C}, often interpreted as continued activity from  the central
object beyond the prompt emission phase. In the normal decay phase that follows,  the jet forward shock
 interacts with the external medium.
 At some point, typically $10^4$ $-$ $10^5$\,s after the GRB, the afterglow light curve  steepens again,
 caused by  energy injection from the central object and by the widening of the jet
\citep[see, e.g.,][]{1999ApJ...519L..17S}.

Several studies have investigated the relationship between the energies associated with the prompt and afterglow phases.~\citet{Liang_2007} find a linear correlation between the prompt isotropic energy in gamma rays $E_{\gamma,\mathrm{prompt}}$ and the afterglow isotropic energy in X-rays $E_{X,\mathrm{afterglow}}$, later confirmed by, amongst others,~\citet{Dainotti_2015}. Similarly,~\citet{10.1111/j.1365-2966.2008.14214.x} find a relationship between $E_{\gamma,\mathrm{prompt}}$ and the kinetic energy in the afterglow $E_{k,\mathrm{afterglow}}$ with $E_{k,\mathrm{afterglow}} \propto E^{0.42}_{\gamma,\mathrm{prompt}}$. This suggests that about 10\% of the blast's kinetic energy is turned into radiation during the prompt phase (see also~\citealt{aksulu_exploring_2022}). Such relationships are crucial for understanding the transition from the prompt to the afterglow. 

ore specifically when looking at the plateau phase,~\citet{10.1111/j.1365-2966.2011.19433.x} find a correlation between
the luminosity at the end of the plateau phase and various parameters related to the energetics and luminosity of the
prompt emission, again substantiating a connection between the prompt and afterglow. Several hypotheses have been
proposed to explain the plateau phase. One such hypothesis involves delayed energy injection from, for example, a
fast-spinning magnetar (e.g.,~\citealt{Zhang_2001,10.1093/mnras/stu1277}). In {another scenario, the plateau
phase is linked to additional energy transfer from slower ejecta hitting the blast wave
(e.g.,~\citealt{10.1093/mnras/stu1921}). 
A model proposed by \citet{2020MNRAS.492.2847B} produces the plateaus for GRBs whose structured jets are viewed off-axis. 
Precession of the GRB jets can also influence the afterglow light curve, and produce a plateau \citep{Huang_2021}.}
ost of these models signify a direct connection between the prompt and afterglow emission, encouraging the research that we describe in the following sections. 

In summary, the combination of
a single-source engine driving both time series, 
and the promising phenomenology of the prompt and afterglow light curves,
motivates us to investigate if a transformer model can connect these two data streams,
as a start to finding the shared underlying physics. {We note that recurrent neural networks have also been used extensively for these kinds of problems but these can struggle to maintain information across longer sequences (although mitigable, too, see ~\citealt{2014arXiv1409.3215S}). The natural ability of transformer models to capture long-range dependencies combined with their efficient processing of larger datasets, makes them an ideal choice for our study.}

\section{Data acquisition}
\label{sec:Transformer_Data}
The Neil Gehrels Swift Observatory, Swift hereafter, is a multi-wavelength observatory built to study GRBs and their
afterglows \citep{2004ApJ...611.1005G}. It is equipped with three main instruments:
the Burst Alert Telescope (BAT; \citealt{2005SSRv..120..143B}), the
X-ray Telescope (XRT; \citealt{2005SSRv..120..165B}), and the Ultraviolet/Optical Telescope (UVOT; \citealt{2005SSRv..120...95R}). The BAT detects GRBs and computes their positions to bring the GRB location within the field of view of the XRT and UVOT. This coordinated response allows Swift to observe the prompt and afterglow emissions in X-ray, ultraviolet, and optical wavelengths, enabling the first large dataset of GRBs with afterglow measurements to be collected~(e.g.,~\citealt{2016ApJ...829....7L}). 

Here we are interested in GRBs with both BAT and XRT measurements. In particular, the Swift Burst Analyser~\citep{2010A&A...519A.102E} provides flux light curves for both instruments in specified energy regimes. In contrast to the count rate data, spectral evolution is taken into account and the data is extrapolated to the desired energy range for a more straightforward comparison between the instruments. In our case, we use the BAT and XRT unabsorbed flux density light curves at 10 keV which is right in between the energy range of the XRT and BAT instruments where little spectral extrapolation is needed to compute the light curves.

We use the Swift/BAT Gamma-Ray Burst Catalog\footnote{\url{https://swift.gsfc.nasa.gov/results/batgrbcat/}} \citep{2016ApJ...829....7L} in
combination with the \texttt{swifttools} Python package\footnote{\url{https://www.swift.ac.uk/API/ukssdc/}} to get the
data of GRBs analysed by the Burst Analyser. For the BAT data, there are multiple binning options available whereas the
XRT data has only one binning option. To retain some freedom in the length of the light curves we give to our neural
networks, we choose the relatively small bin size of 64 ms for the BAT data instead of one of the signal-to-noise (S/N) threshold binning options and rebin later. We do not include bins flagged as "bad" where the uncertainty in the counts-to-flux conversion is too large. Because we work in log space, both in time and flux density, we only include bins after the BAT trigger time which is set to $t=0$.

{Our dataset contains 1132 BAT-detected GRBs which have associated XRT measurements}. The number of points per light curve varies greatly, with some BAT light curves only having a single datapoint and others having more than $10^4$ flux values. One option to obtain equal-length light curves as input for our machine-learning models is zero padding. As the computational complexity of the standard transformer model scales with the length of the sequence squared, it becomes intractable to pad every light curve to the length of the longest BAT GRB light curve with $\sim$ $10^4$ points. Rather, we rebin the BAT data with 500 equally spaced bins between the time of the first and last BAT measurement in our dataset. The XRT light curves generally consist of many fewer data points. Thus, we rebin these light curves with 50 equally spaced bins instead, which is close to the median number of points per XRT light curve. {The heterogenity in the number of measurements for each GRB in the dataset still requires the use of zero padding afterwards; see Sect.~\ref{sec:Transformer_NN} for the technical details and Sect.~\ref{sec:Transformer_disc} for a discussion on the possible biases this introduces.}
\section{Transformer}
\label{sec:Transformer_NN}
\begin{figure}
    \centering
    \includegraphics[width=0.4\textwidth]{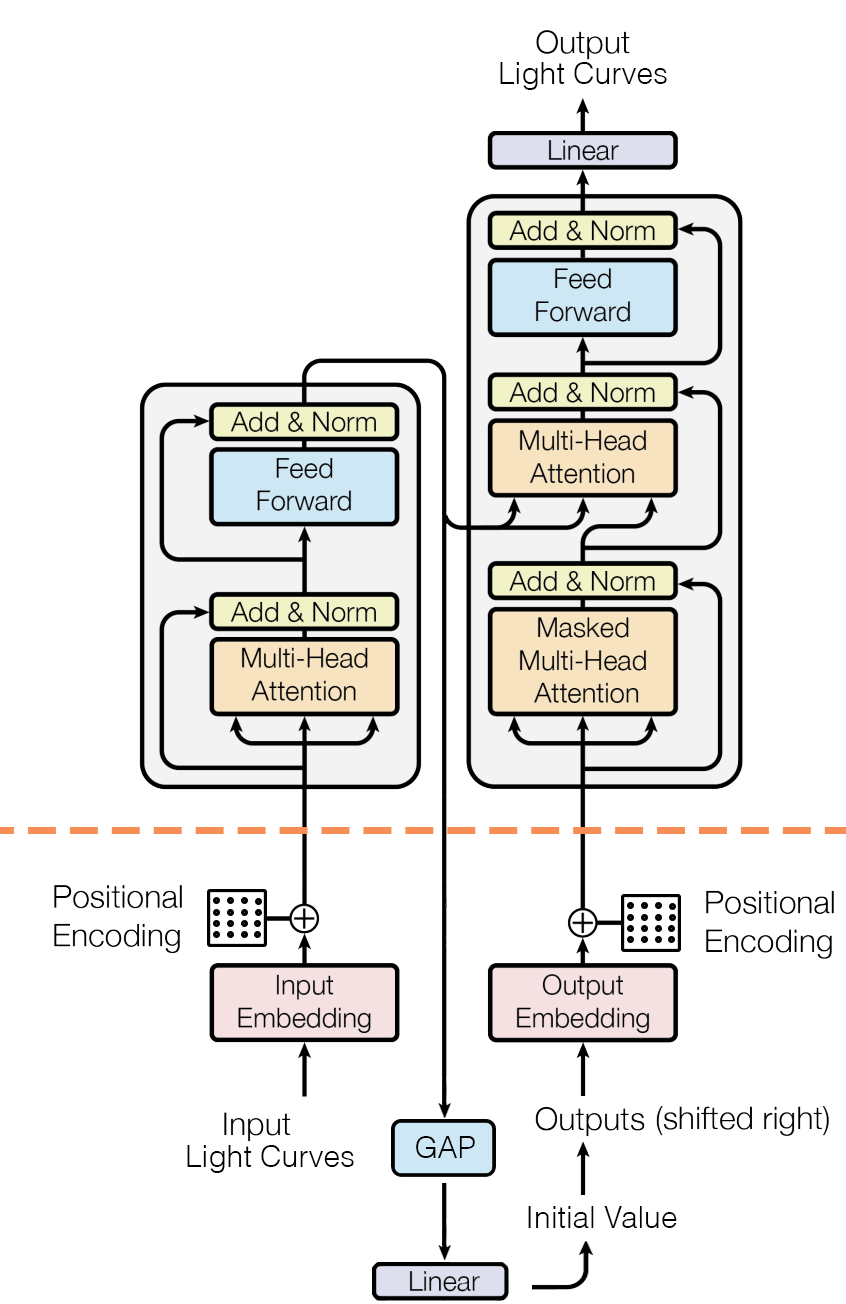}
    \caption{A schematic representation of our transformer network, adapted from~\citet{NIPS2017_3f5ee243}. The modifications made in this work are shown below the orange dashed line.
}
    \label{fig:Transformer_schematic}
\end{figure}
{In this section, we first introduce the basic architecture, and next detail our implementation where relevant. It is thus styled from more general to more specific.}

Our deep-learning model of choice is the transformer model~\citep{NIPS2017_3f5ee243}. After the data is embedded in a numeric representation, the transformer model takes in this sequence of data, like a sentence in English or a BAT light curve, and produces a new sequence, e.g., the translated sentence to Spanish or the XRT light curve. The full transformer architecture consists of two main components: an encoder and a decoder. This structure was first applied to sequence models like LSTM networks~\citep{2014arXiv1409.3215S}. The encoder processes the input sequence, BAT GRB light curves in our case, to extract and compile the features and patterns in the data. The decoder takes this information from the encoder and produces the output sequence, i.e., the XRT light curves.
We summarise the embeddings, the attention mechanism, the encoder, and the decoder below. We refer to~\citet{2021arXiv210506178A} for a detailed, pedagogical explanation of applying transformer {encoder} models to astronomical time-series data. {Note that we, in the current study, add a decoder step, and go from a discriminative  to a generative model for our sequence-to-sequence data. These steps are further detailed below.}

\subsection{Embedding}
In NLP tasks, some kind of embedding is required to convert the input and output words or tokens of the data into a dense vector of continuous numbers suitable for the transformer model. As we are already working with continuous time-series data, we use a simple time-distributed linear projection of our 1D (univariate) light curves to the dimensions of our model $d_\mathrm{model}$. This projection adjusts the light curve data to match the expected input size of the model. We still need to introduce some kind of positional knowledge, i.e., where in the light curves the different data points are located, into the model as this is not directly taken care of by the model architecture like in RNNs. \citet{NIPS2017_3f5ee243} employ fixed positional encodings generated by sinusoidal functions but note that learned positional encodings give nearly identical results. We use the latter and convert the index positions of the data points in our light curves to $d_\mathrm{model}$ and add these projections to the linear projections mentioned above. We thus stick to the positional encoding of the indices of the data points. More sophisticated (observing) time representations like \texttt{time2vec}~\citep{2019arXiv190705321M} {may improve the performance of the model, particularly as GRB prompt and afterglow light curves cover many orders of magnitude in time but can still exhibit short-scale time variability. This is harder to capture using positional encodings. Please see Sect.~\ref{sec:Transformer_disc} for further discussion on this topic}.
\subsection{Attention mechanism}
The core concept of the transformer model is the use of attention without incorporating any recurrence in the model. While the general concept of attention was introduced earlier~\citep{2014arXiv1409.0473B}, the scaled-dot product attention introduced in~\citet{NIPS2017_3f5ee243} has proven highly useful. Given three matrices consisting of sets of queries $Q$, keys $K$, and values $V$, for each query the transformer model looks at all the keys to decide which parts of the data, i.e., the values, to focus on. This is done by computing a dot product between the query and the keys. The outcome represents how much each key aligns with the query. These scores are then used to take a weighted sum of the values. In many applications of the transformer model, the keys and values represent the exact same input, or output, embeddings.

athematically, the scaled dot-product attention is given by:
\begin{equation}
    \mathrm{Attention(Q,K,V)} = \mathrm{softmax}\Big(\frac{QK^{T}}{\sqrt{d_k}}\Big)V.
\end{equation}
First, the dot-product is computed between the query and all keys, and the softmax function converts the dot products into probabilities or weights. A scaling is implemented by the dimension of the keys $d_k$. If $d_k$ is large, this scaling counteracts extremely small gradients in the softmax function for large dot products. The output of the softmax is used as weights that represent the importance of the values relative to each query. In multi-headed attention, the queries, keys, and values are linearly projected $h$ times with different learned projection weights $W^{Q}_i$, $W^{K}_i$, and $W^{V}_i$, {initialised in a standard way~\citep{pmlr-v9-glorot10a}}. The attention is calculated in parallel for each of the $h$ projections and combined afterwards. This enables the model to use multiple specialised attention mechanisms in parallel, each focusing on different patterns or relationships within the light curves. In practice, the queries, keys, and values are derived from the embedding vectors ({with dimension $d_\mathrm{model} = 500$}) of the input BAT or output XRT light curves depending on the type of attention, {either self-attention or cross-attention}, required. {This works in a similar way to NLP applications of the transformer model, where instead of each token being represented by an embedding vector, each data point in our light curves is represented by an embedding vector.}

\subsection{Encoder and decoder}
\label{sec:encoder_and_decoder}
{The encoder processes  input BAT GRB light curves and  this information is fed to the decoder to produce output XRT light curves.}
The encoder consists of two blocks: a multi-headed self-attention block, and a fully connected feed-forward network (FFN). In the self-attention block, different positions in the BAT light curves, or more precisely their embedding vectors, are compared between one another to compute a weighted representation of the input. The output of the self-attention is then fed into the FFN which consists of two linear layers with a ReLU activation function~\citep{glorot2011deep} in between. The output of the FFN is used as input for the decoder. For convergence, residual connections are used around both blocks as well as layer normalisation after each block.

The decoder is similar in setup to the encoder with a few key differences. In the inference stage, predictions for the XRT light curves are generated autoregressively. {This means it makes predictions one step at a time, using the sequence of previously predicted data points to forecast the next point in the sequence.} The data points of the light curve are thus predicted step-by-step where all the data points generated until index $i-1$ are used to predict the data point at index $i$. 

{Transformer models allow for the simultaneous processing of entire sequences. This capability enhances training efficiency by eliminating the need for step-by-step prediction during this phase, an important benefit of transformer models over RNNs. As a result, the entire light curves can be trained on directly.} Care must be taken, however, to not make predictions based on future data points which are visible to the transformer model during training but not during inference. To {maintain the integrity of the autoregressive predictions}, first, the target data, meaning the output, of the decoder is shifted one index to the right so that the decoder is trained to predict the next data point based on the previous ones. Subsequently, the self-attention block uses a "causal mask" such that for data points at index $i$ only the relation to points at an index less than $i$ is computed. {This approach restricts the model's attention to only consider data points preceding the current point being predicted.} 

In between the masked self-attention block and the FFN, there is a cross-attention block that computes the multi-headed attention on the output of the encoder. Thus, the keys and values are the encoder outputs while the queries are the output of the self-attention block. That is to say that every index in the XRT light curve attends, or is related, to all indices in the BAT light curve. 

After the FFN, which is the same as the encoder but with different weights, we use a linear layer to project the output back from $d_\mathrm{model}$ to our 1D XRT light curve data. Again, as in the encoder, residual connections and layer normalisations are added to the decoder.

\subsection{Dense neural network modification}
The decoder needs to be initialised with some input to begin its predictions. In language translation tasks, this is often done by adding a start token to the vocabulary of words that is embedded in some numeric representation. Because we work with continuous numbers directly, it is less trivial to provide a start token not already present in the data. Furthermore, in our testing, we observed that the choice of starting input can significantly influence the quality of the predictions, a point we will return to in Sect.~\ref{sec:Transformer_disc}.  As to not introduce a bias by manually choosing the initial values of the predictions, we modify the model to include a dense neural network that predicts the initial value based on the output of the encoder. It consists of a 1D global average pooling layer operating on the output of the encoder followed by a linear projection to the first value of the XRT light curve. This is not the first value of the final predictions but is related to the start codon used in the NLP domain.

The entire transformer model used in this work is shown schematically in Fig.~\ref{fig:Transformer_schematic} with light curves as input and output, learned positional encoding instead of sinusoidal encoding, linear instead of softmax activation for the final layer, and a dense neural network to initialise the predictions.

\subsection{Setup }
\begin{figure}
    \centering
    \includegraphics[width=\columnwidth]{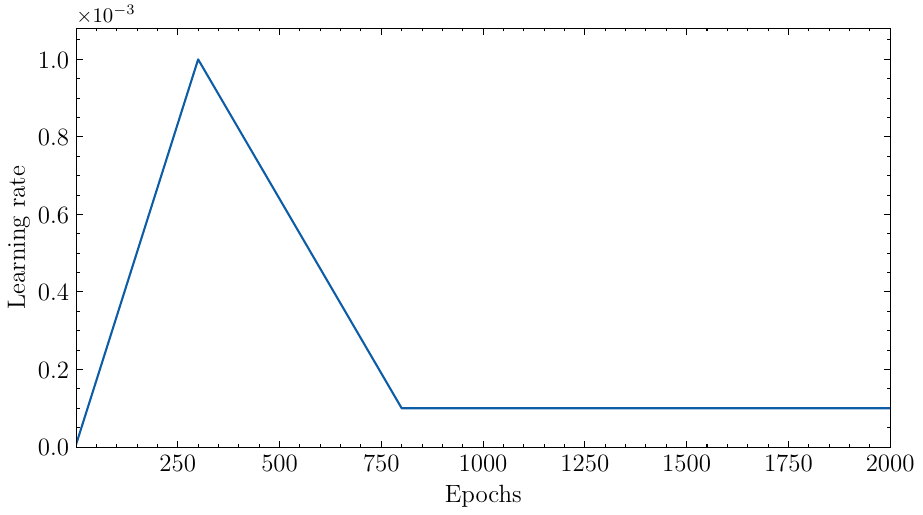}
    \caption{The learning-rate schedule used to train the transformer model.}
    \label{fig:Transformer_lrschedule}
\end{figure}
\begin{figure*}
    \centering
    \includegraphics[width=\columnwidth]{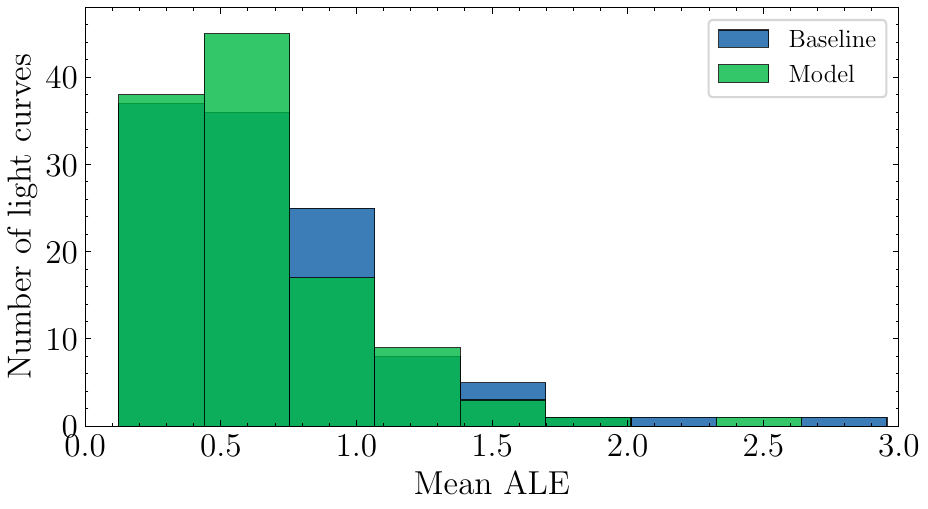}
    \includegraphics[width=\columnwidth]{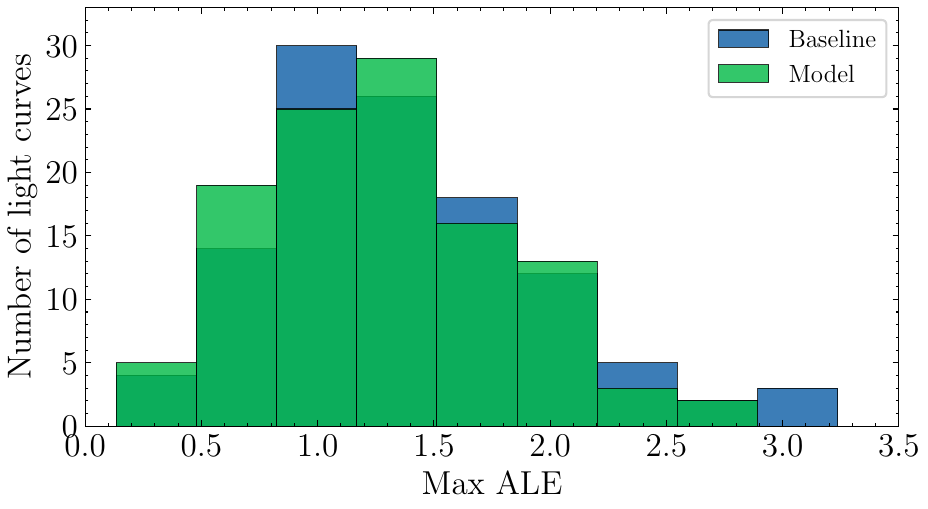}
    \caption{Distributions of the mean and the max ALE for our baseline model and transformer model in the left and right panels, respectively.}
    \label{fig:Transformer_statdist}
\end{figure*}
\subsubsection{Preprocessing}
For preprocessing of the data, we first take the $\log_{10}$ of the flux densities for both the input BAT and output XRT
light curves. We remove the mean per time bin for all bins to get the flux densities in every bin on a similar scale. In
this way, we make sure there is not one bin that will dominate the loss function with very large average flux
densities. In contrast to regular standard scaling (as done in, e.g., \citealt{2022arXiv221210943B}), we do not divide by the standard deviation per bin to preserve differences in variability between bins in hopes that this will aid the model in light curve reconstruction. {Still we apply the normalisation by the mean as mentioned, to make sure outliers do not have too large of an influence on the model performance}. We use an $81/10/9 \ \%$ split for the training, test, and validation dataset size, respectively, {corresponding to 917, 113, and 102 GRBs with associated BAT and XRT light curves}.

any time bins in our dataset have no (i.e., zero) flux, because of missing measurements and zero padding. The model incorporates the masking of zero flux values when calculating the attention scores and the loss function. {Note that while the causal mask (see Sect.~\ref{sec:encoder_and_decoder}) is an inherent part of the self-attention block of the decoder, we require the above-mentioned additional mask to take care of the zero flux values in our data.}  {Thus, these masked values are taken into account too when calculating the initial value of the decoder using the global average pooling layer.} As in~\citet{2022arXiv221210943B}, we use the mean absolute error (MAE) as our loss function, thus modified to disregard the zero flux values. 

\subsubsection{Training}
The learning rate is a key hyperparameter in neural networks, significantly influencing the dynamics of how the network weights are updated during the training process. Adaptive adjustment of the learning rate during training, often referred to as learning rate scheduling, has been shown to enhance performance in many cases (e.g.,~\citealt{2019arXiv190806477W}). Learning-rate scheduling is an important part of training standard transformer models.  Depending on where the layer normalizations are placed, either between the residual blocks as originally proposed (Post-LN transformer model) or inside of the residual connections (Pre-LN transformer model), a learning rate warm-up stage is necessary (Post-LN) or not (Pre-LN)~\citep{xiong2020layer}. This is to avoid instabilities caused by large gradients near the output layers at the initialisation of training. We use the Post-LN transformer model in this work and employ a learning rate schedule where the learning rate is gradually increased from $6\cdot10^{-6}$ to $10^{-3}$ for the first 300 epochs and then slowly decreased for 500 epochs back down to $10^{-4}$ after which it remains constant for the rest of the training. {The learning rate is, in this work, not influenced by the value of the loss function.} The learning rate schedule is visualised in Fig.~\ref{fig:Transformer_lrschedule}. We train our model with the Adam optimiser~\citep{2014arXiv1412.6980K} for 2000 epochs with a batch size of 128 and save the model with the lowest validation loss. 

The transformer model is designed such that both the encoder and decoder blocks can be separately stacked on top of each other to increase the model complexity. Because the size of our training dataset is still rather limited, we stick to one encoder and one decoder block. For the hyperparameters of our model, we set the general dimension $d_\mathrm{model}$ to 500 while the first layer of the FFN has 128 units. Moreover, we use four parallel attention heads in the attention blocks. These parameters were found to produce the best results in a random search of the hyperparameter space given our computing memory limitations. Still, differences in performance were small between sets of hyperparameters. For example, setting $d_\mathrm{model}$ to a smaller value produced similar results. {Our model is fully implemented from scratch in Tensorflow 2.10\footnote{\url{https://www.tensorflow.org/}}, thus no pretrained model was used. Please see Sect.~\ref{sec:Transformer_disc} on the potential benefits of using a pretrained model in future work.}
\section{Results}
\label{sec:Transformer_results}
As a %
baseline model, we take the average, in log space, of the XRT  light curves in our training dataset. 
This baseline model is not as naive as it may appear: the dataset consists of light curves that were all detected with the same instrument, and  we express these in terms of fluence. As the sample is fluence limited,  the largest number of detected light curves will have fluences just above the instrument sensitivity threshold. The average light curve can thus be expected to match most individual light curves quite well.
If we are able to outperform this baseline model {with no dependence on the BAT light curves}, this gives at least some indication that the transformer model is using information from the prompt data to reconstruct the afterglow data. {This serves as additional motivation to utilise this baseline model.}

To quantify our error, we calculate what we will refer to as the Absolute Logarithmic Error (ALE) being the difference in flux between the measured XRT flux densities $\mathrm{F}_\mathrm{observed}$ and the predicted XRT flux densities $\mathrm{F}_\mathrm{pred}$ in log space:

\begin{equation}
    \mathrm{ALE} = |\log_{10}{\mathrm{F}_\mathrm{pred}} - \log_{10}{\mathrm{F}_\mathrm{observed}}|
\end{equation}

We calculate both the mean and the maximum of the ALE per light curve. We show the distributions of the mean and the maximum ALE in the test dataset for both the baseline model and the transformer model in Fig.~\ref{fig:Transformer_statdist}. We also calculate the median of both distributions, thus the median mean ALE and median max ALE, respectively.

\begin{figure}
    \centering
    \includegraphics[width=\columnwidth]{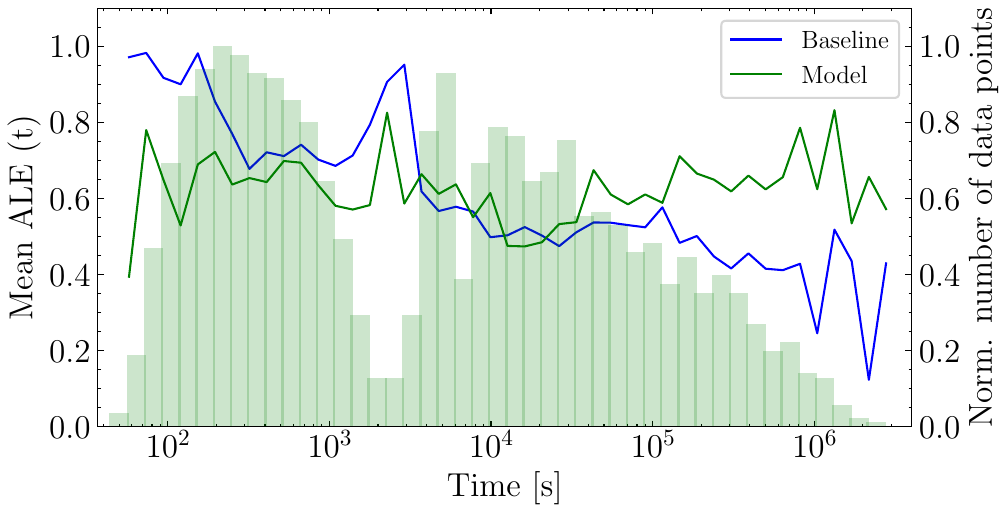}
    \caption{Performance of the baseline model and the transformer model versus time. The mean ALE error is marked on 
    the left y-axis, where lower is better. Shown on the right y-axis is the normalised number of non-zero observations per time bin for which this ALE is calculated.}
    \label{fig:Transformer_overtimemeanfe}
\end{figure}

We are marginally outperforming the baseline model when looking at the median mean ALE which is 0.54 and 0.59 for the transformer model and the baseline, respectively. For the median max ALE, the transformer model is again slightly better with a median of 1.29 vs a median of 1.34 for the baseline model. In Fig.~\ref{fig:Transformer_overtimemeanfe}, we show the mean ALE(t) as a function of time, thus calculated per time bin instead of per light curve. Intuitively, one might expect the prompt emission to have a stronger correlation with the afterglow at earlier times, comparable to the duration of GRBs, than at later ones. This could make it easier for the transformer model to reconstruct the afterglow light curves just after the prompt emission has ended. Furthermore, and perhaps of bigger influence, at early times the transformer model is less reliant on previous predictions to make the next one which could reduce the error, see Sect.~\ref{sec:Transformer_disc}. 

We indeed find we are outperforming the baseline model for times $\lesssim 10^4$ s. We are worse than the baseline model after $10^4$ s, however, possibly because of the smaller number of XRT observations, and thus training data, at those times. The ALE metric is less robust after $\gtrsim 10^6$s as well because of the small number of samples for which it is calculated. 

\begin{figure}
    \hspace{-0.8em}
    \includegraphics[width=\columnwidth]{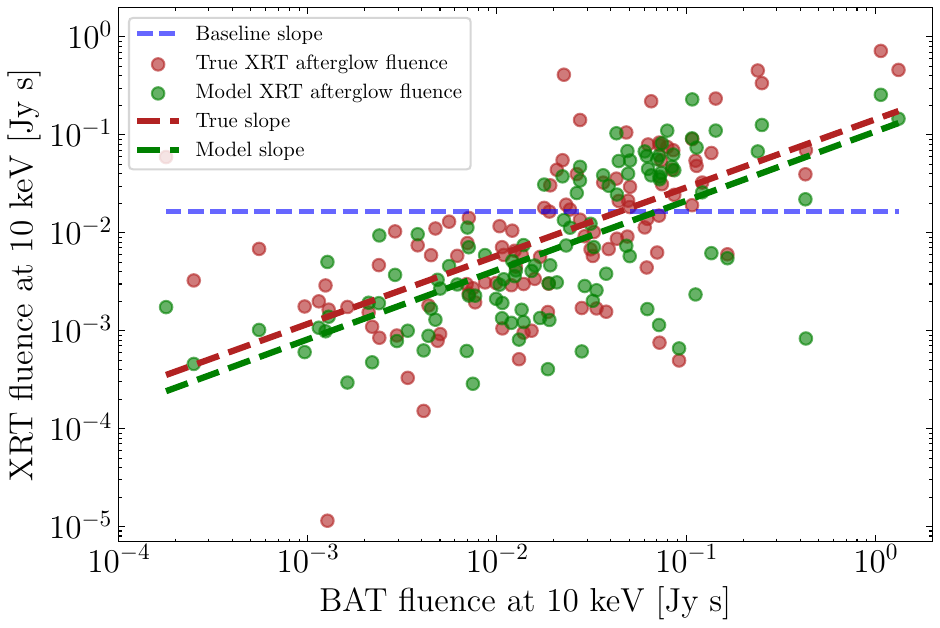}
    \caption{Correlation between the BAT and XRT fluence in both the observed XRT light curves as well as the
    reconstructed XRT light curves. The baseline model (blue) by definition cannot
    reproduce this trend; the model (green) agrees with the data very well.
} 
    \label{fig:Transformer_fluencecorr}
\end{figure}

Around 45 - 60 minutes after the BAT trigger time, a gap is visible in the number of observations. This corresponds to the nominal time allocated for regular GRB observations with Swift as it is limited by Earth occultations\footnote{\href{https://swift.gsfc.nasa.gov/proposals/tech_appd/swiftta_v17.pdf}{The Neil Gehrels Swift Observatory Technical Handbook}}. Both the transformer model and the baseline model show a peak in the ALE around $\sim10^4$s as a result of the dearth of observations around that time. 

any studies have observed correlations between the fluence of the prompt and the afterglow emission~(e.g.,~\citealt{2009MNRAS.397.1177E,10.1093/mnras/sts066}). 
The fact that an equivalent correlation also exists for the isotropic prompt and afterglow energies of GRBs 
with known redshifts \citep{2007ApJ...662.1093W}
shows that this fluence correlation is not just the result of the generally unknown distance to the GRB 
equally affecting the fluence of both phases, and introducing a correlation that way.
We calculate the fluence at 10 keV for both the BAT and XRT data to see if the predictions by our transformer model follow the correlation visible in the observed data. We estimate the fluence by performing a direct numerical integration over the flux densities in each time bin. We show the fluences for our test dataset in Fig.~\ref{fig:Transformer_fluencecorr}. Note that this is specifically the fluence at 10 keV and not over an energy band as is common. Also shown is a linear fit in log space for both the predicted and measured fluences. The best estimates of the slope for the predicted fluences compared to the observed fluences match very well, 0.71 versus 0.69. A positive correlation between the BAT fluence and the XRT fluence of the predictions is clearly visible, just as in the measured fluences. The transformer model is thus able to learn that stronger bursts usually have brighter afterglows. This is somewhat surprising given the fact that, on average, per light curve, the transformer model is not significantly outperforming the baseline model which has slope zero in the fluence-fluence space, the blue dashed line in Fig.~\ref{fig:Transformer_fluencecorr}. Because the transformer model is better at predicting the flux for times $\lesssim 10^4$ s, where most of the fluence comes from, we hypothesise that this is likely why the correlations match well.

\subsection{Test cases}
\begin{figure*}
    \centering
    \includegraphics[width=\columnwidth]{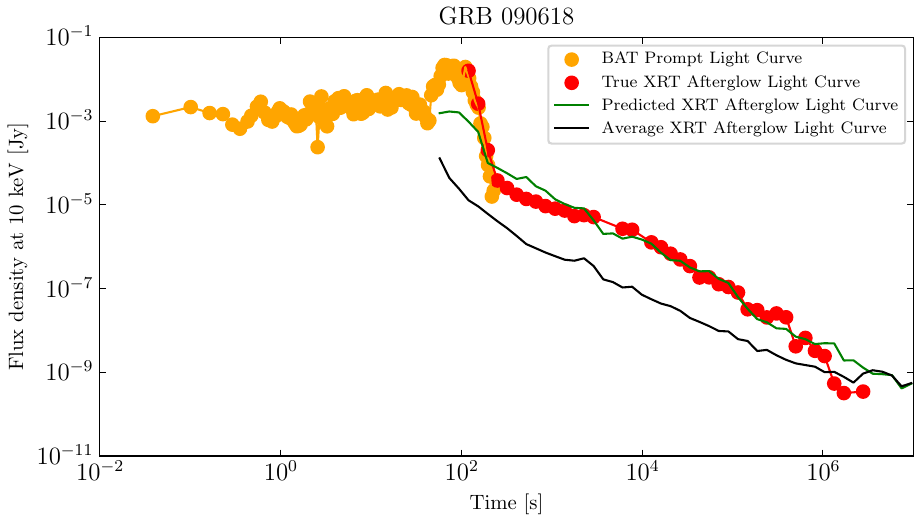}
    \includegraphics[width=\columnwidth]{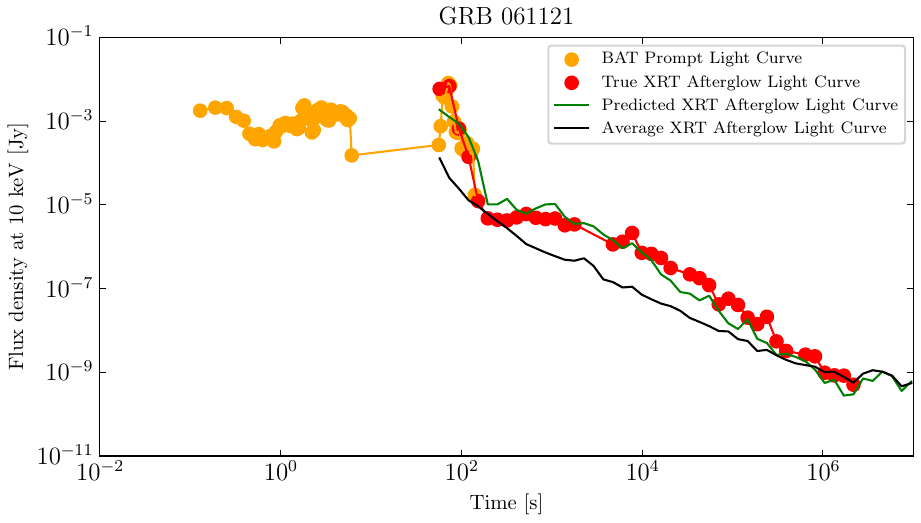}
    \caption{Two examples of reconstructed XRT light curves. The yellow dots show the BAT light curve while the red dots show the XRT light curve. The solid black line shows the average of the XRT light curves in the training dataset. The solid green line shows our reconstruction.}
    \label{fig:Transformer_recon_1}
\end{figure*}

\begin{figure*}
    \centering
    \includegraphics[width=\columnwidth]{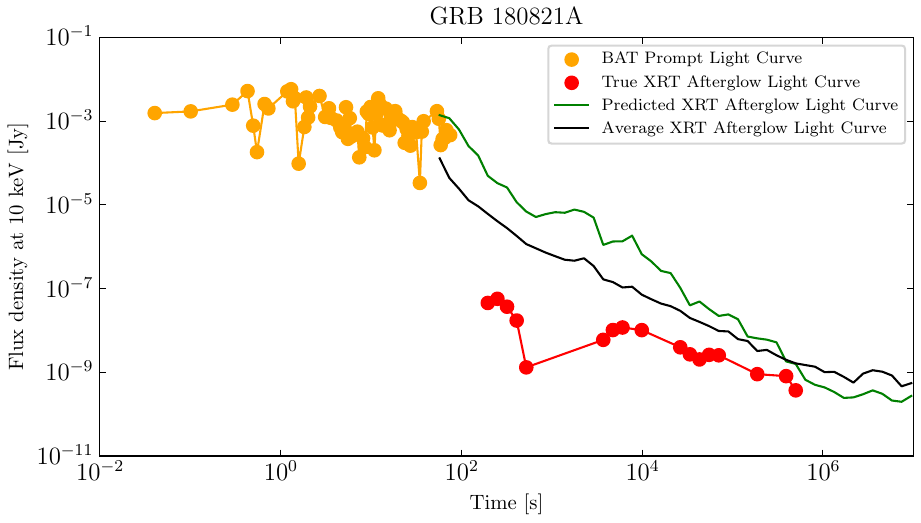}
    \includegraphics[width=\columnwidth]{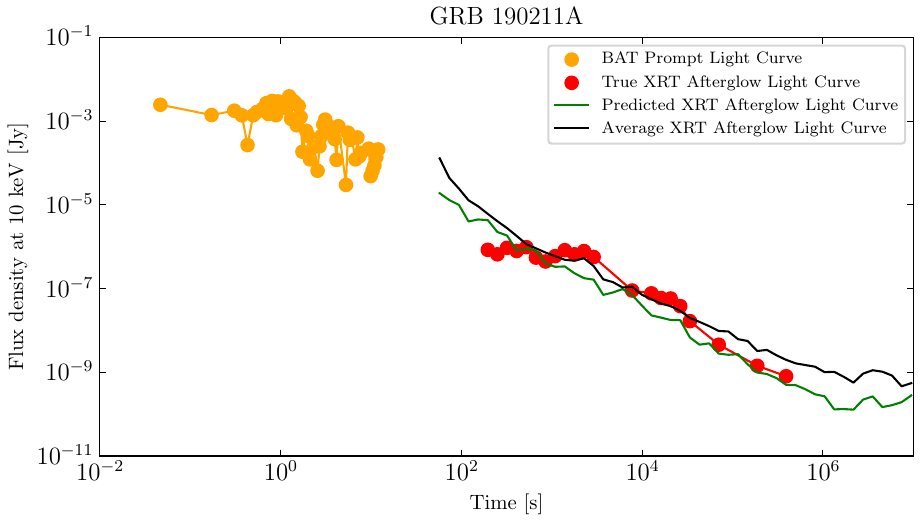}
    \caption{Two further examples of reconstructed XRT light curves. Colours match those of Fig.~\ref{fig:Transformer_recon_1}.}
    \label{fig:Transformer_recon_2}
\end{figure*}

We show representative examples of reconstructed light curves in Fig.~\ref{fig:Transformer_recon_1} and
Fig.~\ref{fig:Transformer_recon_2}.
These evenly include fits from throughout the mean ALE distribution. 
The overall shape and amplitude are well reconstructed for the afterglow light
curve of GRB 090618 \citep{2009GCN..9512....1S}. The predicted afterglow light curve follows a broken power law with a  smooth variation in decay slope~\citep{10.1111/j.1365-2966.2011.19183.x}, in reasonable accordance with the true measurements. GRB 090618 was the brightest burst detected by the XRT at the time of detection and has a canonical afterglow morphology~\citep{2009MNRAS.397.1177E}. Although the high-latitude emission, which is still part of the prompt phase, is underestimated by the reconstruction, the salient features of the afterglow including the plateau, and decay, though not the post-jet break, are present. The mean ALE is 0.28 compared to a mean ALE of 1.1 for the baseline model.


\begin{figure*}
    \centering
    \includegraphics[width=\columnwidth]{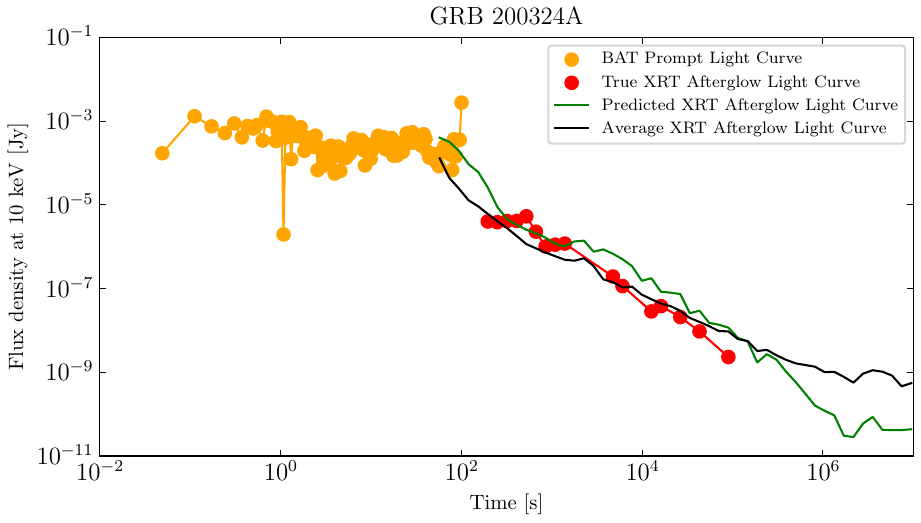}
    \includegraphics[width=\columnwidth]{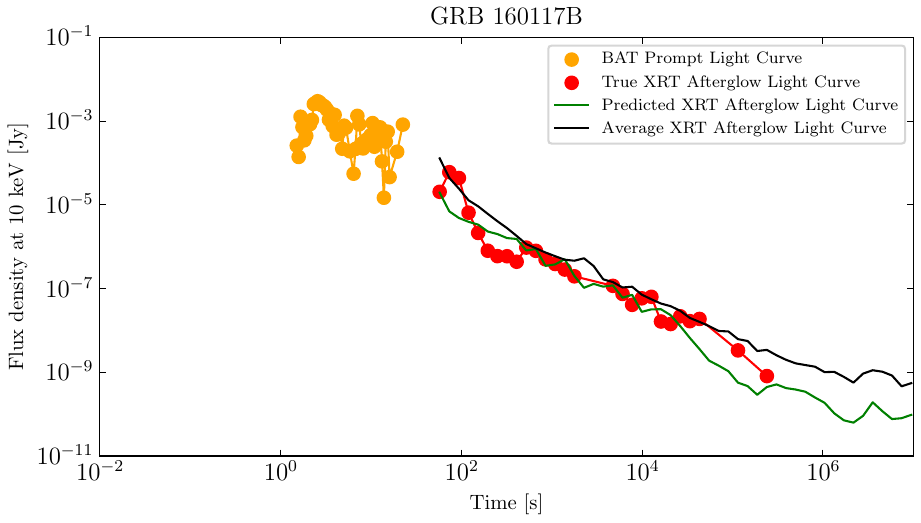}
    \caption{Another two examples of reconstructed XRT light curves. Colours match those of Fig.~\ref{fig:Transformer_recon_1}.}
    \label{fig:Transformer_recon_3}
\end{figure*}

\begin{figure*}
    \centering
    \includegraphics[width=0.975\columnwidth]{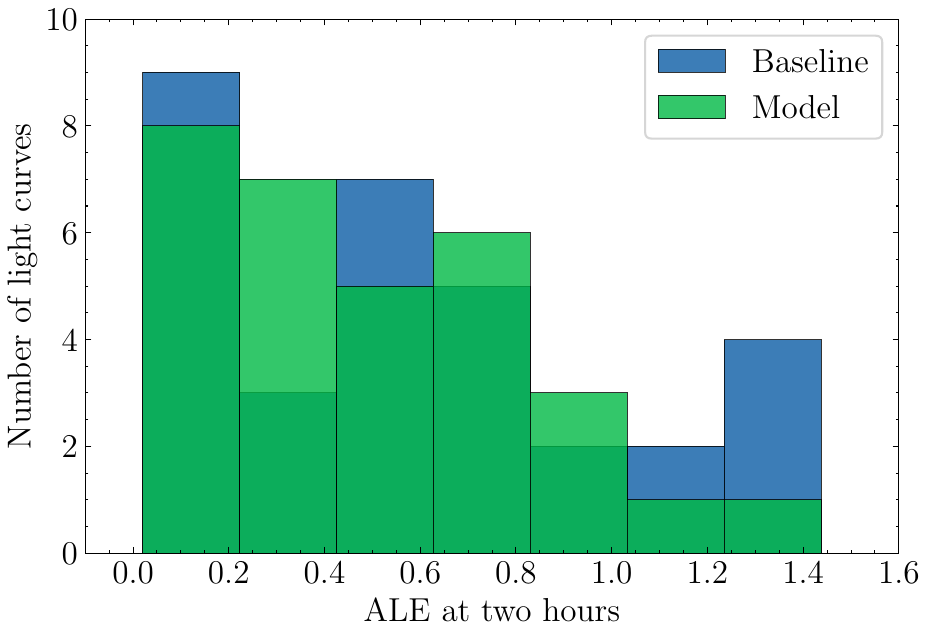}
    \includegraphics[width=\columnwidth]{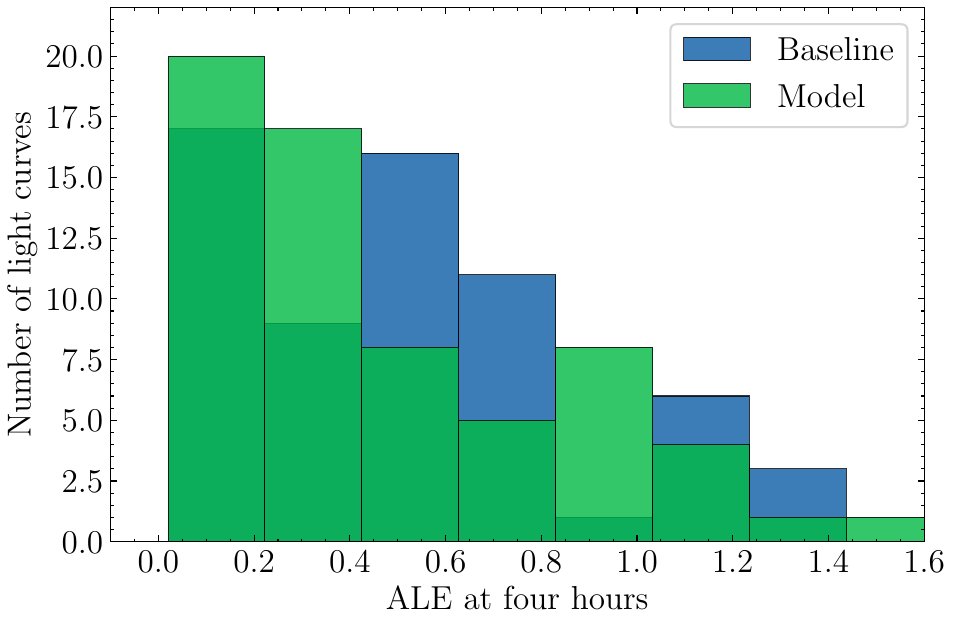}
    \caption{Distributions of the ALE for our baseline model (blue) and transformer model (green) at two hours in the left panel and four hours in the right panel.}
    \label{fig:Transformer_extradist}
\end{figure*}

For GRB 061121 \citep{2006GCN..5823....1P}, the distinct phases of the canonical afterglow light curve are better apparent still in the transformer models reconstruction, despite the slightly higher mean ALE of 0.31. The start and end break times of the plateau phase match, visually, well with the observations as does the high-latitude emission decay slope. The two flares before $10^2$ s and other deviations from the broken power law shape are not convincingly modelled by the transformer model.

For GRB 180821A \citep{2018GCN.23158....1T}, the transformer model reconstruction with a mean ALE of 1.84 is bad and does not outperform the baseline model, which has a, still poor, ALE of 1.22. The reconstruction by the transformer model predicts an afterglow with a much higher fluence than the actual observations suggest. The S/N-binned BAT light curve for this GRB in the Swift Burst Analyser catalogue shows only a few time bins reaching an S/N threshold of 4, and this could be one of the reasons why the transformer model has difficulty translating the BAT data to the XRT data for this GRB. For GRB 090618, for example, there are many bins that reach the S/N 4 threshold in the catalogue. Even so, other GRBs with high S/N BAT measurements in our dataset do not necessarily have good XRT light curve reconstructions.  

The residual errors are low for the reconstructed afterglows of GRB 190211A, GRB 200324A, and GRB 160117B
(\citealt{2019GCN.23883....1M,2020GCN.27428....1D} and \citealt{2016GCN.18875....1S}, respectively),
see Fig.~\ref{fig:Transformer_recon_2} and Fig.~\ref{fig:Transformer_recon_3}, with mean residuals of 0.33, 0.38, and 0.32. The baseline model performs admirably as well, however, with mean residuals of 0.27, 0.26, and 0.34. While the baseline model is naive with no dependence on the prompt emission, quite often it is still an adequate representation of the afterglows in the XRT data and, as such, not easy to outperform with prompt-dependent models like our transformer model.

\subsection{Potential application}
A potential application of a flux predictor like the transformer model in this work is to guide observations of the afterglow based on the measured prompt emission. One of the aims of the Athena mission~\citep{2020AN....341..224B}, for example, is to observe the missing baryons in the warm-hot intergalactic medium (WHIM). Athena will search for signatures of the missing baryons by measuring absorption features in GRB afterglow spectra imprinted by the WHIM~\citep{2020A&A...642A..24W}. Athena's observing strategy is typically not centred on transient events like GRB afterglows, which are fleeting and require rapid response. Ideally, it should be on target after two to four hours for detection of the absorption features to be possible if the afterglow is sufficiently bright~\citep{2020A&A...642A..24W}. Athena needs to be selective in which afterglows to follow up, and it would therefore be beneficial to have a good prediction of the brightness of the afterglow after a few hours based on the prompt emission. Based on Fig.~\ref{fig:Transformer_overtimemeanfe}, we are not yet able to beat the baseline model in predicting the flux density in the two to four hour range. In Fig.~\ref{fig:Transformer_extradist}, we also show the distribution of the ALE at two and four hours. The standard deviation or spread in the ALE is similar between the baseline model and the transformer model. Thus, our transformer model currently does not provide much benefit for guiding such targeted observations. 

\section{Discussion}
\label{sec:Transformer_disc}
{The key finding of the current study is the following. 
The transformer neural network predicts afterglow light curves well for some cases, 
but is generally unable to outperform the baseline model, which has no dependence on specific BAT prompt measurements. While we are able to obtain the expected positive correlation between the prompt and afterglow fluence in the reconstructed dataset, we cannot consistently make better predictions than the baseline model for the afterglow emission using the transformer model.}  

Our exploration of this method has brought to light a number of factors that currently hamper {robust} reconstruction of afterglow light curves using a transformer model. 

\subsection{Challenges in training a sequence-to-sequence model}
\begin{figure}
    \centering
    \includegraphics[width=\columnwidth]{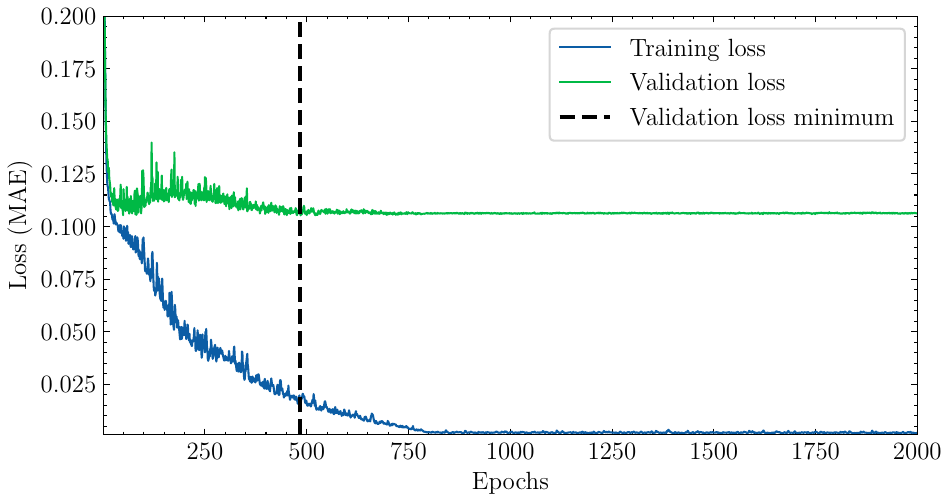}
    \caption{Evolution of the MAE loss during training. The training loss is shown in blue while the validation loss is shown in green. The vertical dashed black line indicates the epoch with the lowest validation loss.}
    \label{fig:Transformer_loss_history}
\end{figure}

\begin{figure}
    \centering
    \vspace{-1em}
    \includegraphics[width=\columnwidth]{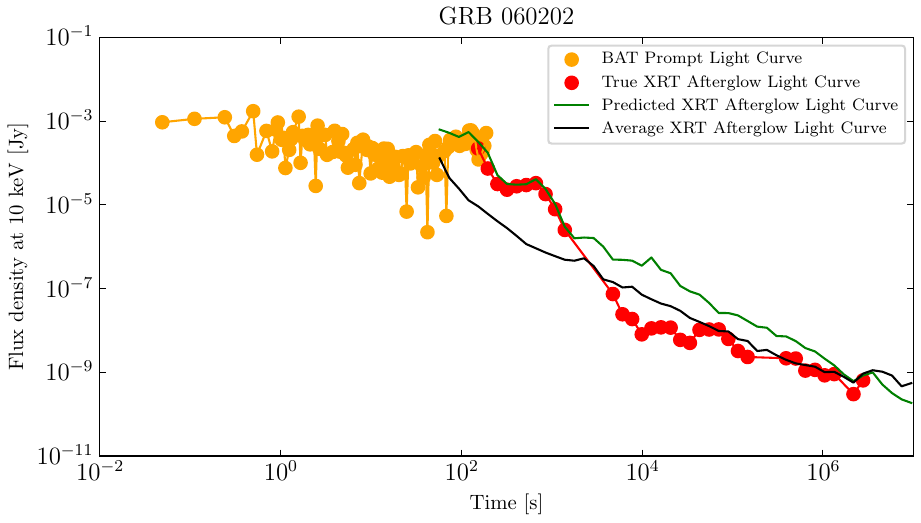}
    \caption{The reconstruction of the light curve of GRB 060202 in the training data set with an overfitted transformer model. Colours match those of Fig.~\ref{fig:Transformer_recon_1}.}
    \label{fig:Transformer_train_recon}
\end{figure}

We have had difficulty in training a model which overfits the training data in the same way a simple dense neural network could. A small two-layer dense network with, for example, 128 units per layer, is quickly able to precisely memorise specific instances in the training data. While this is generally not desired, the ability to overfit a model gives an indication of its ability to learn from the data. We did a separate training run where we saved both the model with the lowest validation loss and the model at the end of training, see Fig.~\ref{fig:Transformer_loss_history}. The median mean ALE on the training data is 0.28 and 0.25, for the validation model and the final model, respectively. While the residual errors are thus lower on the training dataset than on the test dataset, as expected based on the loss values, we do not find a major difference in accuracy between the validation model and the final model. This is surprising given the fact that the training loss is about eight times lower for the model that is trained for 2000 epochs. We find that some light curves in the training data are reconstructed almost perfectly while the model performs worse than the baseline model for other light curves. Furthermore, the reconstruction goes astray for some GRBs right after the initial observation time of 45$-$60 minutes, discussed earlier, where there is a dearth of measurements; see Fig.~\ref{fig:Transformer_train_recon} for example, though this is not necessarily the norm. One specific lesson learned thus is that, to confidently have a transformer model follow the evolution of the afterglow, the dataset should avoid significant gaps but provide steady and dependable sampling. The XRT dataset would be more valuable still if the scheduling could fill in the gap around 60 minutes. 

A possible explanation for the above-mentioned behaviour is the difference between the training and inference of transformer models. As mentioned in Sect.~\ref{sec:Transformer_NN}, during training a technique called "teacher forcing" is used to mimic the autoregressive way in which inference happens. The next prediction of the XRT light curve is based on the ground truth of earlier data points which may skew the results obtained during the training stage. In the inference stage, the next prediction is based on all previous predictions, which could be substantially different from the ground truth. Any errors that exist for previous predictions are then propagated to future predictions which would make the reconstructed light curve less robust during inference than during training.  This is a well-known problem in sequence-to-sequence models like the transformer model, called "exposure bias"~\citep{2015arXiv151106732R}, and solutions exist to alleviate this problem~\citep{2019arXiv190607651M}, but opinions differ regarding the severity of the problem in NLP-related tasks~\citep{2019arXiv190510617H,2022arXiv220401171A}.

In our case, we believe that the high degree of autocorrelation in both the BAT and XRT dataset likely exacerbates this issue. There are also few GRBs that have flux values in all time bins, which adds an extra degree of difficulty as the transformer model has to bridge those gaps while maintaining good reconstructions afterwards based on the predictions in the gap. This is different compared to the dense neural networks where the input is mapped to all outputs independently, which might aid overfitting. On the other hand, the predictions of the transformer model in the test data set are much better than what we achieved early on with simple dense neural networks, exactly because of the autoregressive nature of transformer models. 

\subsection{Measurement uncertainties and data processing}
In this study, we have not accounted for uncertainties in the flux density measurements of the BAT and XRT data. We essentially give equal weight or importance to every flux density value in our data set, while in reality, some of those values have much lower associated uncertainty than others. Although our model provides a useful preliminary analysis, it is important to note that incorporating error estimates on the data points could lead to more robust results. In future work, the loss function can be modified to include a weighting based on the errors in the XRT data points. This may help the transformer model in distinguishing between variations in the data that are statistically significant and those that are within the measurement error. Furthermore, in cases where the uncertainties associated with the BAT light curves are large, the model's predictions may be less reliable, irrespective of its apparent performance.

A related topic is the binning of our light curves. Applying the same binning to each light curve is likely not the optimal way to present the information to our transformer model. For some GRBs, only a few accurate measurements exist for their prompt or afterglow emission, while others have many more robust measurements than the number of bins used in this work. One option is to bin by S/N, which ensures that every data point is of good quality. This does require a change in the positional embedding, which is currently based on the index positions. The indices do not correspond to the same observing times for S/N-binned light curves. Instead, time embeddings like \texttt{time2vec}~\citep{2019arXiv190705321M} could provide a good alternative in this instance. Our current choice of binning is too coarse to properly expose short-scale time variability like flares in the light curves. The modelling of such variations on top of the general broken power-law shape could benefit from an S/N-binned approach as well. Furthermore, the overall reconstruction of the afterglow time series using transformer models might benefit from additional, distinct features like flares.

The initial input to the decoder is something that merits further exploration. If, for example, the decoder input is initialised, or seeded, with the final flux density value of the prompt emission, this already gives crucial information on the relation between the strength of the prompt emission and the brightness of the afterglow. Here, we chose to base the initial input on the output of the encoder with a straightforward dense neural network. Increasing the complexity of this network could improve this first prediction which, given the problem of exposure bias, might have a significant impact on the overall time series reconstruction ability of the transformer model. We reiterate that this start value is not the first value of the final predictions, and instead represents a value like the start codon used in the NLP domain but tuned towards the different GRB light curves.

\subsection{Physical relations}
{In the introduction we laid out our aims of finding a causal relationship between the prompt and afterglow time series, 
and potentially next determining which underlying joint physics can explain this relation. 
As our transformer model better predicts the afterglow flux for the first hour than the baseline model can, a correlation for that part at least is apparent.
This is, however, a correlation that was noted before ~\citep[e.g.,][]{Gehrels_2008}, and we can offer no insights beyond the existing theories. 
Beyond the hour, the transformer model cannot reproduce the afterglow based on the prompt emission. 
Apparently, the prompt-emission features it was trained on do not contain or present sufficient information for predicting the evolution of the afterglow. 
Overall, we are not yet able to infer new temporal associations between the prompt and afterglow light curves, nor new GRB physics behaviour such as jet structure and evolution, or shocks and medium characteristics.}
\subsection{Future work}
It is clear that, given our current transformer network and the data set that is available for combined BAT and XRT measurements of GRBs, we are not able to consistently predict the afterglow emission from the prompt emission. It is less obvious how to improve the reconstructions. The successful prediction of afterglow emission from prompt emission hinges not only on the quality and volume of the dataset but also on the validity of the hypothesis that such a connection truly exists in nature. We are likely limited by the amount of data at any rate, which is small compared to usual machine learning data sets. However, many GRBs are observed with much finer time resolution than that we chose for this study, which could be leveraged to our advantage. An approach to accessing such information would likely require investigating self-supervised pre-training strategies, which have proven decisive and very fruitful in the NLP domain~\citep{radford2018improving}, followed by fine-tuning to our channel-to-channel 'translation' task, which falls outside of the scope of this work. 

Work must be done to understand how much data is needed to confidently claim or disprove a strong correlation between the prompt and afterglow emission, beyond merely considering summary statistics such as the fluence. A potential approach involves the creation of a substantial synthetic dataset wherein the afterglow emission is derived from the prompt emission through a specific dependence. By incrementally training the transformer model on expanding dataset sizes, either in the number of time series used or in the number of time bins per time series, we could ascertain the point at which the given correlation can be confidently established.

Further establishing what kind of correlation we wish to model, could reduce the degrees of freedom to be constrained. Though, this likely means putting some physical assumptions back into the methodology. For example, focusing purely on predicting the occurrence of flares in the light curves could provide a more targeted approach. If we could reliably predict the flares in the afterglow based on the prompt emission, this would offer compelling evidence that certain short-term physics contributing to the prompt phase is also influential in the generation of the afterglow. Another way to reduce the degrees of freedom is to first fit the afterglow light curves like in~\citet{10.1093/mnras/sts066}, and predict the broken power law slopes and break times instead of the full light curve. It would also be interesting to explore any differences in the reconstruction of long versus short GRB afterglows, although the size of the data set for the latter would again be a limiting factor.

\subsubsection{Pretrained models}
Our dataset contains $>$1000\,GRBs and we attempted to train the transformer model from scratch. 
If a large dataset and succesful model exist that encode (partly) the same physics, 
then these could be the foundation on which to specialise our GRB model.
Building on such foundation models has been very successful in the NLP domain. 
One interesting option for GRBs, are data and models of blazars (Active Galactic Nuclei whose relativistic jets are pointed at Earth).
The energetics of GRBs and blazars have already been shown to follow the same scaling \citep{2012Sci...338.1445N}, and blazars outnumber GRBs by about a factor 5~\citep{2015Ap&SS.357...75M}. Still, that number is most likely not sufficient, and blazars evolve on much longer timescales than GRBs. Reversely, if future GRB missions significantly increase the number of GRB afterglows, and improve their coverage in time, such that a large model can be successfully trained, this could next be applied to other, less common relativistic jets in, for example, microquasars or AGN jets.

{The capacity of the transformer model to abstract and learn from such diverse data through their embedding in a common vector space is one of its key strengths (e.g,~\citealt{jaegle2021perceiver}). Even so, tackling the heterogeneity in the input data, e.g. the differences in the detectors across their various observation bands, is important for training foundation models in a consistent way. There are various strategies to overcome these differences based on data preprocessing, feature engineering, and a variety of other methods (e.g.,~\citealt{hendrycks2019augmix}). For example, metadata about the detectors could be included as part of the input features. This can help the model learn detector-specific biases or sensitivities and adjust its weights accordingly~\citep{nagrani2021attention}. We encourage further work on these topics.}

\section{Summary and Conclusions}
\label{sec:Transformer_conc}
A number of correlations between measured and inferred parameters of the prompt and afterglow phases of GRBs have been established. To what extent the physics of both phases is connected, is not yet known. 
In an attempt to identify new relations between the prompt and afterglow sequences, 
that could subsequently be physically interpreted,  
we have used the transformer model, motivated by its recent success as a sequence-to-sequence
translator. We attempt to directly predict GRB afterglow emission from the associated prompt emission. We used the Swift-BAT and XRT prompt and afterglow data sets to train a small transformer model consisting of one encoder and one decoder. The model was adjusted to include a novel dense neural network setup to predict the initial value of the predictions.
The trained transformer model marginally outperformed the baseline model, i.e., the average of the XRT afterglow light curves, but did not prove to be a consistent improvement over it. Still, we recovered the prompt-afterglow fluence-fluence correlation accurately based on the reconstructed light curves. Furthermore, in some cases, the transformer model managed to realistically predict the different phases of a canonical afterglow, including the plateau phase. 

This work is the first, to our knowledge, to outline a methodology using transformer models for sequence-to-sequence prediction applied to astrophysical data. This methodology can potentially establish a causal connection at the population level between two time-resolved components in different wavebands (or counterparts) of any type of astrophysical transient. Future work on including observational uncertainties,  improved binning, and exploring how much additional data is required, may lead to finding various connections between the prompt and afterglow phases of GRBs, but at the moment transformer models are not yet able to identify such new correlations. %

\section*{Acknowledgements}
We thank Ralph Wijers for fruitful discussions and a careful reading of this manuscript and Maarten Stol for stimulating dialog.

This research was supported by the Dutch Research Council (NWO) 
through Vici research program 'ARGO' with project number 639.043.815 and through CORTEX (NWA.1160.18.316), under the research programme NWA-ORC. OMB also acknowledges funding from the Leids Kerkhoven-Bosscha Fonds (LKBF).

\section*{Data and Code Availability}
Between acceptance and publication the code for data acquisition and analysis
will be made available at: \url{http://doi.org/10.5281/zenodo.10932872}

\bibliographystyle{rasti}
\bibliography{references} %

\begin{thebibliography}{77}
\expandafter\ifx\csname natexlab\endcsname\relax\def\natexlab#1{#1}\fi

\bibitem[Aksulu et~al.(2022)Aksulu, Wijers, van Eerten, \& van~der
  Horst]{aksulu_exploring_2022}
Aksulu, M.~D., Wijers, R. A. M.~J., van Eerten, H.~J., \& van~der Horst, A.~J.,
  2022.
\newblock Exploring the {GRB} population: robust afterglow modelling, {\it
  Monthly Notices of the Royal Astronomical Society\/}, {\bf 511}(2),
  2848--2867.

\bibitem[{Allam} \& {McEwen}(2021)]{2021arXiv210506178A}
{Allam}, Tarek, J. \& {McEwen}, J.~D., 2021.
\newblock {Paying Attention to Astronomical Transients: Introducing the
  Time-series Transformer for Photometric Classification}, {\it arXiv
  e-prints\/}.

\bibitem[{Arora} et~al.(2022){Arora}, {El Asri}, {Bahuleyan}, \& {Kit
  Cheung}]{2022arXiv220401171A}
{Arora}, K., {El Asri}, L., {Bahuleyan}, H., \& {Kit Cheung}, J.~C., 2022.
\newblock {Why Exposure Bias Matters: An Imitation Learning Perspective of
  Error Accumulation in Language Generation}, {\it arXiv e-prints\/}.

\bibitem[{Bahdanau} et~al.(2014){Bahdanau}, {Cho}, \&
  {Bengio}]{2014arXiv1409.0473B}
{Bahdanau}, D., {Cho}, K., \& {Bengio}, Y., 2014.
\newblock {Neural Machine Translation by Jointly Learning to Align and
  Translate}, {\it arXiv e-prints\/}.

\bibitem[{Barret} et~al.(2020){Barret}, {Decourchelle}, {Fabian}, {Guainazzi},
  {Nandra}, {Smith}, \& {den Herder}]{2020AN....341..224B}
{Barret}, D., {Decourchelle}, A., {Fabian}, A., {Guainazzi}, M., {Nandra}, K.,
  {Smith}, R., \& {den Herder}, J.-W., 2020.
\newblock {The Athena space X-ray observatory and the astrophysics of hot
  plasma}, {\it Astronomische Nachrichten\/}, {\bf 341}(2), 224--235.

\bibitem[{Barthelmy} et~al.(2005){Barthelmy}, {Barbier}, {Cummings},
  {Fenimore}, {Gehrels}, {Hullinger}, {Krimm}, {Markwardt}, {Palmer},
  {Parsons}, {Sato}, {Suzuki}, {Takahashi}, {Tashiro}, \&
  {Tueller}]{2005SSRv..120..143B}
{Barthelmy}, S.~D., {Barbier}, L.~M., {Cummings}, J.~R., {Fenimore}, E.~E.,
  {Gehrels}, N., {Hullinger}, D., {Krimm}, H.~A., {Markwardt}, C.~B., {Palmer},
  D.~M., {Parsons}, A., {Sato}, G., {Suzuki}, M., {Takahashi}, T., {Tashiro},
  M., \& {Tueller}, J., 2005.
\newblock {The Burst Alert Telescope (BAT) on the SWIFT Midex Mission}, {\it
  Space Science Reviews\/}, {\bf 120}(3-4), 143--164.

\bibitem[{Beniamini} et~al.(2020){Beniamini}, {Duque}, {Daigne}, \&
  {Mochkovitch}]{2020MNRAS.492.2847B}
{Beniamini}, P., {Duque}, R., {Daigne}, F., \& {Mochkovitch}, R., 2020.
\newblock {X-ray plateaus in gamma-ray bursts' light curves from jets viewed
  slightly off-axis}, {\it \mnras\/}, {\bf 492}(2), 2847--2857.

\bibitem[Berger(2014)]{berger2014short}
Berger, E., 2014.
\newblock Short-duration gamma-ray bursts, {\it Annual review of Astronomy and
  Astrophysics\/}, {\bf 52}, 43--105.

\bibitem[Bernardini(2015)]{BERNARDINI201564}
Bernardini, M.~G., 2015.
\newblock Gamma-ray bursts and magnetars: Observational signatures and
  predictions, {\it Journal of High Energy Astrophysics\/}, {\bf 7}, 64--72.

\bibitem[{Boersma} \& {van Leeuwen}(2023)]{2022arXiv221210943B}
{Boersma}, O.~M. \& {van Leeuwen}, J., 2023.
\newblock {DeepGlow: An efficient neural network emulator of physical afterglow
  models for gamma-ray bursts and gravitational-wave events}, {\it \pasa\/},
  {\bf 40}, e030.

\bibitem[Brown et~al.(2020)Brown, Mann, Ryder, Subbiah, Kaplan, Dhariwal,
  Neelakantan, Shyam, Sastry, Askell, Agarwal, Herbert-Voss, Krueger, Henighan,
  Child, Ramesh, Ziegler, Wu, Winter, Hesse, Chen, Sigler, Litwin, Gray, Chess,
  Clark, Berner, McCandlish, Radford, Sutskever, \& Amodei]{GPT3}
Brown, T., Mann, B., Ryder, N., Subbiah, M., Kaplan, J.~D., Dhariwal, P.,
  Neelakantan, A., Shyam, P., Sastry, G., Askell, A., Agarwal, S.,
  Herbert-Voss, A., Krueger, G., Henighan, T., Child, R., Ramesh, A., Ziegler,
  D., Wu, J., Winter, C., Hesse, C., Chen, M., Sigler, E., Litwin, M., Gray,
  S., Chess, B., Clark, J., Berner, C., McCandlish, S., Radford, A., Sutskever,
  I., \& Amodei, D., 2020.
\newblock Language models are few-shot learners, in {\em Advances in Neural
  Information Processing Systems\/}, vol.~33, pp. 1877--1901, Curran
  Associates, Inc.

\bibitem[{Burrows} et~al.(2005){Burrows}, {Hill}, {Nousek}, {Kennea}, {Wells},
  {Osborne}, {Abbey}, {Beardmore}, {Mukerjee}, {Short}, {Chincarini},
  {Campana}, {Citterio}, {Moretti}, {Pagani}, {Tagliaferri}, {Giommi},
  {Capalbi}, {Tamburelli}, {Angelini}, {Cusumano}, {Br{\"a}uninger}, {Burkert},
  \& {Hartner}]{2005SSRv..120..165B}
{Burrows}, D.~N., {Hill}, J.~E., {Nousek}, J.~A., {Kennea}, J.~A., {Wells}, A.,
  {Osborne}, J.~P., {Abbey}, A.~F., {Beardmore}, A., {Mukerjee}, K., {Short},
  A.~D.~T., {Chincarini}, G., {Campana}, S., {Citterio}, O., {Moretti}, A.,
  {Pagani}, C., {Tagliaferri}, G., {Giommi}, P., {Capalbi}, M., {Tamburelli},
  F., {Angelini}, L., {Cusumano}, G., {Br{\"a}uninger}, H.~W., {Burkert}, W.,
  \& {Hartner}, G.~D., 2005.
\newblock {The Swift X-Ray Telescope}, {\it Space Science Reviews\/}, {\bf
  120}(3-4), 165--195.

\bibitem[{Campana} et~al.(2005){Campana}, {Antonelli}, {Chincarini}, {Covino},
  {Cusumano}, {Malesani}, {Mangano}, {Moretti}, {Pagani}, {Romano},
  {Tagliaferri}, {Capalbi}, {Perri}, {Giommi}, {Angelini}, {Boyd}, {Burrows},
  {Hill}, {Gronwall}, {Kennea}, {Kobayashi}, {Kumar}, {M{\'e}sz{\'a}ros},
  {Nousek}, {Roming}, {Zhang}, {Abbey}, {Beardmore}, {Breeveld}, {Goad},
  {Godet}, {Mason}, {Osborne}, {Page}, {Poole}, \&
  {Gehrels}]{2005ApJ...625L..23C}
{Campana}, S., {Antonelli}, L.~A., {Chincarini}, G., {Covino}, S., {Cusumano},
  G., {Malesani}, D., {Mangano}, V., {Moretti}, A., {Pagani}, C., {Romano}, P.,
  {Tagliaferri}, G., {Capalbi}, M., {Perri}, M., {Giommi}, P., {Angelini}, L.,
  {Boyd}, P., {Burrows}, D.~N., {Hill}, J.~E., {Gronwall}, C., {Kennea}, J.~A.,
  {Kobayashi}, S., {Kumar}, P., {M{\'e}sz{\'a}ros}, P., {Nousek}, J.~A.,
  {Roming}, P.~W.~A., {Zhang}, B., {Abbey}, A.~F., {Beardmore}, A.~P.,
  {Breeveld}, A., {Goad}, M.~R., {Godet}, O., {Mason}, K.~O., {Osborne}, J.~P.,
  {Page}, K.~L., {Poole}, T., \& {Gehrels}, N., 2005.
\newblock {Swift Observations of GRB 050128: The Early X-Ray Afterglow}, {\it
  The Astrophysical Journal\/}, {\bf 625}(1), L23--L26.

\bibitem[{Connor} \& {van Leeuwen}(2018)]{2018AJ....156..256C}
{Connor}, L. \& {van Leeuwen}, J., 2018.
\newblock {Applying Deep Learning to Fast Radio Burst Classification}, {\it The
  Astronomical Journal\/}, {\bf 156}(6), 256.

\bibitem[Dainotti \& {Del Vecchio}(2017)]{DAINOTTI201723}
Dainotti, M. \& {Del Vecchio}, R., 2017.
\newblock Gamma ray burst afterglow and prompt-afterglow relations: An
  overview, {\it New Astronomy Reviews\/}, {\bf 77}, 23--61.

\bibitem[Dainotti et~al.(2011)Dainotti, Ostrowski, \&
  Willingale]{10.1111/j.1365-2966.2011.19433.x}
Dainotti, M.~G., Ostrowski, M., \& Willingale, R., 2011.
\newblock {Towards a standard gamma-ray burst: tight correlations between the
  prompt and the afterglow plateau phase emission}, {\it Monthly Notices of the
  Royal Astronomical Society\/}, {\bf 418}(4), 2202--2206.

\bibitem[Dainotti et~al.(2015)Dainotti, Vecchio, Shigehiro, \&
  Capozziello]{Dainotti_2015}
Dainotti, M.~G., Vecchio, R.~D., Shigehiro, N., \& Capozziello, S., 2015.
\newblock Selection effects in gamma-ray burst correlations: Consequences on
  the ratio between gamma-ray burst and star formation rates, {\it The
  Astrophysical Journal\/}, {\bf 800}(1), 31.

\bibitem[Davies et~al.(2019)Davies, Serjeant, \&
  Bromley]{10.1093/mnras/stz1288}
Davies, A., Serjeant, S., \& Bromley, J.~M., 2019.
\newblock {Using convolutional neural networks to identify gravitational lenses
  in astronomical images}, {\it Monthly Notices of the Royal Astronomical
  Society\/}, {\bf 487}(4), 5263--5271.

\bibitem[{Dichiara} et~al.(2020){Dichiara}, {Barthelmy}, {Beardmore},
  {Bernardini}, {D'Avanzo}, {D'Elia}, {Gropp}, {Kennea}, {Klingler}, {Laha},
  {Marshall}, {Melandri}, {Page}, {Palmer}, {Sbarrato}, {Sbarufatti},
  {Ukwatta}, \& {Neil Gehrels Swift Observatory Team}]{2020GCN.27428....1D}
{Dichiara}, S., {Barthelmy}, S.~D., {Beardmore}, A.~P., {Bernardini}, M.~G.,
  {D'Avanzo}, P., {D'Elia}, V., {Gropp}, J.~D., {Kennea}, J.~A., {Klingler},
  N.~J., {Laha}, S., {Marshall}, F.~E., {Melandri}, A., {Page}, K.~L.,
  {Palmer}, D.~M., {Sbarrato}, T., {Sbarufatti}, B., {Ukwatta}, T.~N., \& {Neil
  Gehrels Swift Observatory Team}, 2020.
\newblock {GRB 200324A: Swift detection of a burst}, {\it GRB Coordinates
  Network\/}, {\bf 27428}, 1.

\bibitem[{Dosovitskiy} et~al.(2020){Dosovitskiy}, {Beyer}, {Kolesnikov},
  {Weissenborn}, {Zhai}, {Unterthiner}, {Dehghani}, {Minderer}, {Heigold},
  {Gelly}, {Uszkoreit}, \& {Houlsby}]{2020arXiv201011929D}
{Dosovitskiy}, A., {Beyer}, L., {Kolesnikov}, A., {Weissenborn}, D., {Zhai},
  X., {Unterthiner}, T., {Dehghani}, M., {Minderer}, M., {Heigold}, G.,
  {Gelly}, S., {Uszkoreit}, J., \& {Houlsby}, N., 2020.
\newblock {An Image is Worth 16x16 Words: Transformers for Image Recognition at
  Scale}, {\it arXiv e-prints\/}.

\bibitem[{Evans} et~al.(2009){Evans}, {Beardmore}, {Page}, {Osborne},
  {O'Brien}, {Willingale}, {Starling}, {Burrows}, {Godet}, {Vetere}, {Racusin},
  {Goad}, {Wiersema}, {Angelini}, {Capalbi}, {Chincarini}, {Gehrels}, {Kennea},
  {Margutti}, {Morris}, {Mountford}, {Pagani}, {Perri}, {Romano}, \&
  {Tanvir}]{2009MNRAS.397.1177E}
{Evans}, P.~A., {Beardmore}, A.~P., {Page}, K.~L., {Osborne}, J.~P., {O'Brien},
  P.~T., {Willingale}, R., {Starling}, R.~L.~C., {Burrows}, D.~N., {Godet}, O.,
  {Vetere}, L., {Racusin}, J., {Goad}, M.~R., {Wiersema}, K., {Angelini}, L.,
  {Capalbi}, M., {Chincarini}, G., {Gehrels}, N., {Kennea}, J.~A., {Margutti},
  R., {Morris}, D.~C., {Mountford}, C.~J., {Pagani}, C., {Perri}, M., {Romano},
  P., \& {Tanvir}, N., 2009.
\newblock {Methods and results of an automatic analysis of a complete sample of
  Swift-XRT observations of GRBs}, {\it Monthly Notices of the Royal
  Astronomical Society\/}, {\bf 397}(3), 1177--1201.

\bibitem[{Evans} et~al.(2010){Evans}, {Willingale}, {Osborne}, {O'Brien},
  {Page}, {Markwardt}, {Barthelmy}, {Beardmore}, {Burrows}, {Pagani},
  {Starling}, {Gehrels}, \& {Romano}]{2010A&A...519A.102E}
{Evans}, P.~A., {Willingale}, R., {Osborne}, J.~P., {O'Brien}, P.~T., {Page},
  K.~L., {Markwardt}, C.~B., {Barthelmy}, S.~D., {Beardmore}, A.~P., {Burrows},
  D.~N., {Pagani}, C., {Starling}, R.~L.~C., {Gehrels}, N., \& {Romano}, P.,
  2010.
\newblock {The Swift Burst Analyser. I. BAT and XRT spectral and flux evolution
  of gamma ray bursts}, {\it Astronomy \& Astrophysics\/}, {\bf 519}, A102.

\bibitem[{Galama} et~al.(1998){Galama}, {Vreeswijk}, {van Paradijs},
  {Kouveliotou}, {Augusteijn}, {B{\"o}hnhardt}, {Brewer}, {Doublier},
  {Gonzalez}, {Leibundgut}, {Lidman}, {Hainaut}, {Patat}, {Heise}, {in't Zand},
  {Hurley}, {Groot}, {Strom}, {Mazzali}, {Iwamoto}, {Nomoto}, {Umeda},
  {Nakamura}, {Young}, {Suzuki}, {Shigeyama}, {Koshut}, {Kippen}, {Robinson},
  {de Wildt}, {Wijers}, {Tanvir}, {Greiner}, {Pian}, {Palazzi}, {Frontera},
  {Masetti}, {Nicastro}, {Feroci}, {Costa}, {Piro}, {Peterson}, {Tinney},
  {Boyle}, {Cannon}, {Stathakis}, {Sadler}, {Begam}, \&
  {Ianna}]{1998Natur.395..670G}
{Galama}, T.~J., {Vreeswijk}, P.~M., {van Paradijs}, J., {Kouveliotou}, C.,
  {Augusteijn}, T., {B{\"o}hnhardt}, H., {Brewer}, J.~P., {Doublier}, V.,
  {Gonzalez}, J.~F., {Leibundgut}, B., {Lidman}, C., {Hainaut}, O.~R., {Patat},
  F., {Heise}, J., {in't Zand}, J., {Hurley}, K., {Groot}, P.~J., {Strom},
  R.~G., {Mazzali}, P.~A., {Iwamoto}, K., {Nomoto}, K., {Umeda}, H.,
  {Nakamura}, T., {Young}, T.~R., {Suzuki}, T., {Shigeyama}, T., {Koshut}, T.,
  {Kippen}, M., {Robinson}, C., {de Wildt}, P., {Wijers}, R.~A.~M.~J.,
  {Tanvir}, N., {Greiner}, J., {Pian}, E., {Palazzi}, E., {Frontera}, F.,
  {Masetti}, N., {Nicastro}, L., {Feroci}, M., {Costa}, E., {Piro}, L.,
  {Peterson}, B.~A., {Tinney}, C., {Boyle}, B., {Cannon}, R., {Stathakis}, R.,
  {Sadler}, E., {Begam}, M.~C., \& {Ianna}, P., 1998.
\newblock {An unusual supernova in the error box of the
  {\ensuremath{\gamma}}-ray burst of 25 April 1998}, {\it Nature\/}, {\bf
  395}(6703), 670--672.

\bibitem[{Gehrels} et~al.(2004){Gehrels}, {Chincarini}, {Giommi}, {Mason},
  {Nousek}, {Wells}, {White}, {Barthelmy}, {Burrows}, {Cominsky}, {Hurley},
  {Marshall}, {M{\'e}sz{\'a}ros}, {Roming}, {Angelini}, {Barbier}, {Belloni},
  {Campana}, {Caraveo}, {Chester}, {Citterio}, {Cline}, {Cropper}, {Cummings},
  {Dean}, {Feigelson}, {Fenimore}, {Frail}, {Fruchter}, {Garmire}, {Gendreau},
  {Ghisellini}, {Greiner}, {Hill}, {Hunsberger}, {Krimm}, {Kulkarni}, {Kumar},
  {Lebrun}, {Lloyd-Ronning}, {Markwardt}, {Mattson}, {Mushotzky}, {Norris},
  {Osborne}, {Paczynski}, {Palmer}, {Park}, {Parsons}, {Paul}, {Rees},
  {Reynolds}, {Rhoads}, {Sasseen}, {Schaefer}, {Short}, {Smale}, {Smith},
  {Stella}, {Tagliaferri}, {Takahashi}, {Tashiro}, {Townsley}, {Tueller},
  {Turner}, {Vietri}, {Voges}, {Ward}, {Willingale}, {Zerbi}, \&
  {Zhang}]{2004ApJ...611.1005G}
{Gehrels}, N., {Chincarini}, G., {Giommi}, P., {Mason}, K.~O., {Nousek}, J.~A.,
  {Wells}, A.~A., {White}, N.~E., {Barthelmy}, S.~D., {Burrows}, D.~N.,
  {Cominsky}, L.~R., {Hurley}, K.~C., {Marshall}, F.~E., {M{\'e}sz{\'a}ros},
  P., {Roming}, P.~W.~A., {Angelini}, L., {Barbier}, L.~M., {Belloni}, T.,
  {Campana}, S., {Caraveo}, P.~A., {Chester}, M.~M., {Citterio}, O., {Cline},
  T.~L., {Cropper}, M.~S., {Cummings}, J.~R., {Dean}, A.~J., {Feigelson},
  E.~D., {Fenimore}, E.~E., {Frail}, D.~A., {Fruchter}, A.~S., {Garmire},
  G.~P., {Gendreau}, K., {Ghisellini}, G., {Greiner}, J., {Hill}, J.~E.,
  {Hunsberger}, S.~D., {Krimm}, H.~A., {Kulkarni}, S.~R., {Kumar}, P.,
  {Lebrun}, F., {Lloyd-Ronning}, N.~M., {Markwardt}, C.~B., {Mattson}, B.~J.,
  {Mushotzky}, R.~F., {Norris}, J.~P., {Osborne}, J., {Paczynski}, B.,
  {Palmer}, D.~M., {Park}, H.~S., {Parsons}, A.~M., {Paul}, J., {Rees}, M.~J.,
  {Reynolds}, C.~S., {Rhoads}, J.~E., {Sasseen}, T.~P., {Schaefer}, B.~E.,
  {Short}, A.~T., {Smale}, A.~P., {Smith}, I.~A., {Stella}, L., {Tagliaferri},
  G., {Takahashi}, T., {Tashiro}, M., {Townsley}, L.~K., {Tueller}, J.,
  {Turner}, M.~J.~L., {Vietri}, M., {Voges}, W., {Ward}, M.~J., {Willingale},
  R., {Zerbi}, F.~M., \& {Zhang}, W.~W., 2004.
\newblock {The Swift Gamma-Ray Burst Mission}, {\it The Astrophysical
  Journal\/}, {\bf 611}(2), 1005--1020.

\bibitem[{Gehrels} et~al.(2006){Gehrels}, {Norris}, {Barthelmy}, {Granot},
  {Kaneko}, {Kouveliotou}, {Markwardt}, {M{\'e}sz{\'a}ros}, {Nakar}, {Nousek},
  {O'Brien}, {Page}, {Palmer}, {Parsons}, {Roming}, {Sakamoto}, {Sarazin},
  {Schady}, {Stamatikos}, \& {Woosley}]{2006Natur.444.1044G}
{Gehrels}, N., {Norris}, J.~P., {Barthelmy}, S.~D., {Granot}, J., {Kaneko}, Y.,
  {Kouveliotou}, C., {Markwardt}, C.~B., {M{\'e}sz{\'a}ros}, P., {Nakar}, E.,
  {Nousek}, J.~A., {O'Brien}, P.~T., {Page}, M., {Palmer}, D.~M., {Parsons},
  A.~M., {Roming}, P.~W.~A., {Sakamoto}, T., {Sarazin}, C.~L., {Schady}, P.,
  {Stamatikos}, M., \& {Woosley}, S.~E., 2006.
\newblock {A new {\ensuremath{\gamma}}-ray burst classification scheme from
  GRB060614}, {\it Nature\/}, {\bf 444}(7122), 1044--1046.

\bibitem[Gehrels et~al.(2008)Gehrels, Barthelmy, Burrows, Cannizzo, Chincarini,
  Fenimore, Kouveliotou, O’Brien, Palmer, Racusin, Roming, Sakamoto, Tueller,
  Wijers, \& Zhang]{Gehrels_2008}
Gehrels, N., Barthelmy, S.~D., Burrows, D.~N., Cannizzo, J.~K., Chincarini, G.,
  Fenimore, E., Kouveliotou, C., O’Brien, P., Palmer, D.~M., Racusin, J.,
  Roming, P. W.~A., Sakamoto, T., Tueller, J., Wijers, R. A. M.~J., \& Zhang,
  B., 2008.
\newblock Correlations of prompt and afterglow emission in swift long and short
  gamma-ray bursts, {\it The Astrophysical Journal\/}, {\bf 689}(2), 1161.

\bibitem[Ghisellini et~al.(2009)Ghisellini, Nardini, Ghirlanda, \&
  Celotti]{10.1111/j.1365-2966.2008.14214.x}
Ghisellini, G., Nardini, M., Ghirlanda, G., \& Celotti, A., 2009.
\newblock {A unifying view of gamma-ray burst afterglows}, {\it Monthly Notices
  of the Royal Astronomical Society\/}, {\bf 393}(1), 253--271.

\bibitem[Glorot \& Bengio(2010)]{pmlr-v9-glorot10a}
Glorot, X. \& Bengio, Y., 2010.
\newblock Understanding the difficulty of training deep feedforward neural
  networks, in {\em Proceedings of the Thirteenth International Conference on
  Artificial Intelligence and Statistics\/}, vol.~9 of {\bf Proceedings of
  Machine Learning Research}, pp. 249--256, PMLR, Chia Laguna Resort, Sardinia,
  Italy.

\bibitem[Glorot et~al.(2011)Glorot, Bordes, \& Bengio]{glorot2011deep}
Glorot, X., Bordes, A., \& Bengio, Y., 2011.
\newblock Deep sparse rectifier neural networks, in {\em Proceedings of the
  fourteenth international conference on artificial intelligence and
  statistics\/}, pp. 315--323, JMLR Workshop and Conference Proceedings.

\bibitem[Gu et~al.(2015)Gu, Wang, Kuen, Ma, Shahroudy, Shuai, Liu, Wang, Wang,
  Cai, \& Chen]{Gu2015RecentAI}
Gu, J., Wang, Z., Kuen, J., Ma, L., Shahroudy, A., Shuai, B., Liu, T., Wang,
  X., Wang, G., Cai, J., \& Chen, T., 2015.
\newblock Recent advances in convolutional neural networks, {\it Pattern
  Recognition\/}, {\bf 77}, 354--377.

\bibitem[{He} et~al.(2019){He}, {Zhang}, {Zhou}, \&
  {Glass}]{2019arXiv190510617H}
{He}, T., {Zhang}, J., {Zhou}, Z., \& {Glass}, J., 2019.
\newblock {Exposure Bias versus Self-Recovery: Are Distortions Really
  Incremental for Autoregressive Text Generation?}, {\it arXiv e-prints\/}.

\bibitem[Hendrycks et~al.(2019)Hendrycks, Mu, Cubuk, Zoph, Gilmer, \&
  Lakshminarayanan]{hendrycks2019augmix}
Hendrycks, D., Mu, N., Cubuk, E.~D., Zoph, B., Gilmer, J., \& Lakshminarayanan,
  B., 2019.
\newblock Augmix: A simple data processing method to improve robustness and
  uncertainty, {\it arXiv preprint arXiv:1912.02781\/}.

\bibitem[Huang \& Liu(2021)]{Huang_2021}
Huang, B.-Q. \& Liu, T., 2021.
\newblock Energy injection driven by precessing jets in gamma-ray burst
  afterglows, {\it The Astrophysical Journal\/}, {\bf 916}(2), 71.

\bibitem[Jaegle et~al.(2021)Jaegle, Gimeno, Brock, Vinyals, Zisserman, \&
  Carreira]{jaegle2021perceiver}
Jaegle, A., Gimeno, F., Brock, A., Vinyals, O., Zisserman, A., \& Carreira, J.,
  2021.
\newblock Perceiver: General perception with iterative attention, in {\em
  International conference on machine learning\/}, pp. 4651--4664, PMLR.

\bibitem[Jia et~al.(2022)Jia, Sun, Li, Song, Ning, Wei, \& Luo]{Jia_2023}
Jia, P., Sun, R., Li, N., Song, Y., Ning, R., Wei, H., \& Luo, R., 2022.
\newblock Detection of strongly lensed arcs in galaxy clusters with
  transformers, {\it The Astronomical Journal\/}, {\bf 165}(1), 26.

\bibitem[{Kingma} \& {Ba}(2014)]{2014arXiv1412.6980K}
{Kingma}, D.~P. \& {Ba}, J., 2014.
\newblock {Adam: A Method for Stochastic Optimization}, {\it arXiv e-prints\/}.

\bibitem[{Kobayashi} et~al.(1997){Kobayashi}, {Piran}, \&
  {Sari}]{1997ApJ...490...92K}
{Kobayashi}, S., {Piran}, T., \& {Sari}, R., 1997.
\newblock {Can Internal Shocks Produce the Variability in Gamma-Ray Bursts?},
  {\it The Astrophysical Journal\/}, {\bf 490}, 92.

\bibitem[{Kouveliotou} et~al.(1993){Kouveliotou}, {Meegan}, {Fishman}, {Bhat},
  {Briggs}, {Koshut}, {Paciesas}, \& {Pendleton}]{1993ApJ...413L.101K}
{Kouveliotou}, C., {Meegan}, C.~A., {Fishman}, G.~J., {Bhat}, N.~P., {Briggs},
  M.~S., {Koshut}, T.~M., {Paciesas}, W.~S., \& {Pendleton}, G.~N., 1993.
\newblock {Identification of Two Classes of Gamma-Ray Bursts}, {\it The
  Astrophysical Journal\/}, {\bf 413}, L101.

\bibitem[Liang \& Zhang(2005)]{Liang_2005}
Liang, E. \& Zhang, B., 2005.
\newblock Model-independent multivariable gamma-ray burst luminosity indicator
  and its possible cosmological implications, {\it The Astrophysical
  Journal\/}, {\bf 633}(2), 611.

\bibitem[Liang et~al.(2007)Liang, Zhang, \& Zhang]{Liang_2007}
Liang, E.-W., Zhang, B.-B., \& Zhang, B., 2007.
\newblock A comprehensive analysis of swift xrt data. ii. diverse physical
  origins of the shallow decay segment, {\it The Astrophysical Journal\/}, {\bf
  670}(1), 565.

\bibitem[{Lien} et~al.(2016){Lien}, {Sakamoto}, {Barthelmy}, {Baumgartner},
  {Cannizzo}, {Chen}, {Collins}, {Cummings}, {Gehrels}, {Krimm}, {Markwardt},
  {Palmer}, {Stamatikos}, {Troja}, \& {Ukwatta}]{2016ApJ...829....7L}
{Lien}, A., {Sakamoto}, T., {Barthelmy}, S.~D., {Baumgartner}, W.~H.,
  {Cannizzo}, J.~K., {Chen}, K., {Collins}, N.~R., {Cummings}, J.~R.,
  {Gehrels}, N., {Krimm}, H.~A., {Markwardt}, C.~B., {Palmer}, D.~M.,
  {Stamatikos}, M., {Troja}, E., \& {Ukwatta}, T.~N., 2016.
\newblock {The Third Swift Burst Alert Telescope Gamma-Ray Burst Catalog}, {\it
  The Astrophysical Journal\/}, {\bf 829}(1), 7.

\bibitem[Liu et~al.(2017)Liu, Gu, \& Zhang]{liu2017neutrino}
Liu, T., Gu, W.-M., \& Zhang, B., 2017.
\newblock Neutrino-dominated accretion flows as the central engine of gamma-ray
  bursts, {\it New Astronomy Reviews\/}, {\bf 79}, 1--25.

\bibitem[Margutti et~al.(2012)Margutti, Zaninoni, Bernardini, Chincarini,
  Pasotti, Guidorzi, Angelini, Burrows, Capalbi, Evans, Gehrels, Kennea,
  Mangano, Moretti, Nousek, Osborne, Page, Perri, Racusin, Romano, Sbarufatti,
  Stafford, \& Stamatikos]{10.1093/mnras/sts066}
Margutti, R., Zaninoni, E., Bernardini, M.~G., Chincarini, G., Pasotti, F.,
  Guidorzi, C., Angelini, L., Burrows, D.~N., Capalbi, M., Evans, P.~A.,
  Gehrels, N., Kennea, J., Mangano, V., Moretti, A., Nousek, J., Osborne,
  J.~P., Page, K.~L., Perri, M., Racusin, J., Romano, P., Sbarufatti, B.,
  Stafford, S., \& Stamatikos, M., 2012.
\newblock {The prompt-afterglow connection in gamma-ray bursts: a comprehensive
  statistical analysis of Swift X-ray light curves}, {\it Monthly Notices of
  the Royal Astronomical Society\/}, {\bf 428}(1), 729--742.

\bibitem[{Marshall} et~al.(2019){Marshall}, {Evans}, {Gropp}, {Kennea},
  {Klingler}, {Krimm}, {Page}, {Palmer}, {Sbarufatti}, {Siegel}, {Tohuvavohu},
  \& {Tooke}]{2019GCN.23883....1M}
{Marshall}, F.~E., {Evans}, P.~A., {Gropp}, J.~D., {Kennea}, J.~A., {Klingler},
  N.~J., {Krimm}, H.~A., {Page}, K.~L., {Palmer}, D.~M., {Sbarufatti}, B.,
  {Siegel}, M.~H., {Tohuvavohu}, A., \& {Tooke}, S.~F., 2019.
\newblock {GRB 190211A: Swift detection of a burst.}, {\it GRB Coordinates
  Network\/}, {\bf 23883}, 1.

\bibitem[{Massaro} et~al.(2015){Massaro}, {Maselli}, {Leto}, {Marchegiani},
  {Perri}, {Giommi}, \& {Piranomonte}]{2015Ap&SS.357...75M}
{Massaro}, E., {Maselli}, A., {Leto}, C., {Marchegiani}, P., {Perri}, M.,
  {Giommi}, P., \& {Piranomonte}, S., 2015.
\newblock {The 5th edition of the Roma-BZCAT. A short presentation}, {\it
  Astrophysics and Space Science\/}, {\bf 357}(1), 75.

\bibitem[{Mehran Kazemi} et~al.(2019){Mehran Kazemi}, {Goel}, {Eghbali},
  {Ramanan}, {Sahota}, {Thakur}, {Wu}, {Smyth}, {Poupart}, \&
  {Brubaker}]{2019arXiv190705321M}
{Mehran Kazemi}, S., {Goel}, R., {Eghbali}, S., {Ramanan}, J., {Sahota}, J.,
  {Thakur}, S., {Wu}, S., {Smyth}, C., {Poupart}, P., \& {Brubaker}, M., 2019.
\newblock {Time2Vec: Learning a Vector Representation of Time}, {\it arXiv
  e-prints\/}.

\bibitem[Meszaros(2002)]{meszaros2002theories}
Meszaros, P., 2002.
\newblock Theories of gamma-ray bursts, {\it Annual Review of Astronomy and
  Astrophysics\/}, {\bf 40}(1), 137--169.

\bibitem[{Mihaylova} \& {Martins}(2019)]{2019arXiv190607651M}
{Mihaylova}, T. \& {Martins}, A. F.~T., 2019.
\newblock {Scheduled Sampling for Transformers}, {\it arXiv e-prints\/}.

\bibitem[Nagrani et~al.(2021)Nagrani, Yang, Arnab, Jansen, Schmid, \&
  Sun]{nagrani2021attention}
Nagrani, A., Yang, S., Arnab, A., Jansen, A., Schmid, C., \& Sun, C., 2021.
\newblock Attention bottlenecks for multimodal fusion, {\it Advances in neural
  information processing systems\/}, {\bf 34}, 14200--14213.

\bibitem[{Nemmen} et~al.(2012){Nemmen}, {Georganopoulos}, {Guiriec}, {Meyer},
  {Gehrels}, \& {Sambruna}]{2012Sci...338.1445N}
{Nemmen}, R.~S., {Georganopoulos}, M., {Guiriec}, S., {Meyer}, E.~T.,
  {Gehrels}, N., \& {Sambruna}, R.~M., 2012.
\newblock {A Universal Scaling for the Energetics of Relativistic Jets from
  Black Hole Systems}, {\it Science\/}, {\bf 338}(6113), 1445.

\bibitem[{Page} et~al.(2006){Page}, {Barthelmy}, {Beardmore}, {Brown},
  {Burrows}, {Evans}, {Gehrels}, {Guidorzi}, {Kennea}, {Krimm}, {Landsman},
  {Mangano}, {Marshall}, {Osborne}, {Pagani}, {Palmer}, {Perri}, {Romano},
  {Sakamoto}, {Stamatikos}, {Starling}, {Troja}, {vanden Berk}, \&
  {Ziaeepour}]{2006GCN..5823....1P}
{Page}, K.~L., {Barthelmy}, S.~D., {Beardmore}, A.~P., {Brown}, P.~J.,
  {Burrows}, D.~N., {Evans}, P.~A., {Gehrels}, N., {Guidorzi}, C., {Kennea},
  J.~A., {Krimm}, H.~A., {Landsman}, W.~L., {Mangano}, V., {Marshall}, F.~E.,
  {Osborne}, J.~P., {Pagani}, C., {Palmer}, D.~M., {Perri}, M., {Romano}, P.,
  {Sakamoto}, T., {Stamatikos}, M., {Starling}, R.~L.~C., {Troja}, E., {vanden
  Berk}, D.~E., \& {Ziaeepour}, H., 2006.
\newblock {GRB 061121: Swift detection of a bright burst with an optical
  counterpart.}, {\it GRB Coordinates Network\/}, {\bf 5823}, 1.

\bibitem[Page et~al.(2011)Page, Starling, Fitzpatrick, Pandey, Osborne, Schady,
  McBreen, Campana, Ukwatta, Pagani, Beardmore, \&
  Evans]{10.1111/j.1365-2966.2011.19183.x}
Page, K.~L., Starling, R. L.~C., Fitzpatrick, G., Pandey, S.~B., Osborne,
  J.~P., Schady, P., McBreen, S., Campana, S., Ukwatta, T.~N., Pagani, C.,
  Beardmore, A.~P., \& Evans, P.~A., 2011.
\newblock {GRB 090618: detection of thermal X-ray emission from a bright
  gamma-ray burst}, {\it Monthly Notices of the Royal Astronomical Society\/},
  {\bf 416}(3), 2078--2089.

\bibitem[Popham et~al.(1999)Popham, Woosley, \&
  Fryer]{popham1999hyperaccreting}
Popham, R., Woosley, S.~E., \& Fryer, C., 1999.
\newblock Hyperaccreting black holes and gamma-ray bursts, {\it The
  Astrophysical Journal\/}, {\bf 518}(1), 356.

\bibitem[Radford et~al.(2018)Radford, Narasimhan, Salimans, Sutskever,
  et~al.]{radford2018improving}
Radford, A., Narasimhan, K., Salimans, T., Sutskever, I., et~al., 2018.
\newblock Improving language understanding by generative pre-training.

\bibitem[{Ranzato} et~al.(2015){Ranzato}, {Chopra}, {Auli}, \&
  {Zaremba}]{2015arXiv151106732R}
{Ranzato}, M., {Chopra}, S., {Auli}, M., \& {Zaremba}, W., 2015.
\newblock {Sequence Level Training with Recurrent Neural Networks}, {\it arXiv
  e-prints\/}.

\bibitem[Rao et~al.(2021)Rao, Meier, Sercu, Ovchinnikov, \&
  Rives]{rao2021transformer}
Rao, R., Meier, J., Sercu, T., Ovchinnikov, S., \& Rives, A., 2021.
\newblock Transformer protein language models are unsupervised structure
  learners, in {\em International Conference on Learning Representations\/}.

\bibitem[{Rees} \& {M\'esz\'aros}(1992)]{1992MNRAS.258P..41R}
{Rees}, M.~J. \& {M\'esz\'aros}, P., 1992.
\newblock {Relativistic fireballs - Energy conversion and time-scales}, {\it
  Monthly Notices of the Royal Astronomical Society\/}, {\bf 258}, 41.

\bibitem[{Roming} et~al.(2005){Roming}, {Kennedy}, {Mason}, {Nousek}, {Ahr},
  {Bingham}, {Broos}, {Carter}, {Hancock}, {Huckle}, {Hunsberger}, {Kawakami},
  {Killough}, {Koch}, {McLelland}, {Smith}, {Smith}, {Soto}, {Boyd},
  {Breeveld}, {Holland}, {Ivanushkina}, {Pryzby}, {Still}, \&
  {Stock}]{2005SSRv..120...95R}
{Roming}, P. W.~A., {Kennedy}, T.~E., {Mason}, K.~O., {Nousek}, J.~A., {Ahr},
  L., {Bingham}, R.~E., {Broos}, P.~S., {Carter}, M.~J., {Hancock}, B.~K.,
  {Huckle}, H.~E., {Hunsberger}, S.~D., {Kawakami}, H., {Killough}, R., {Koch},
  T.~S., {McLelland}, M.~K., {Smith}, K., {Smith}, P.~J., {Soto}, J.~C.,
  {Boyd}, P.~T., {Breeveld}, A.~A., {Holland}, S.~T., {Ivanushkina}, M.,
  {Pryzby}, M.~S., {Still}, M.~D., \& {Stock}, J., 2005.
\newblock {The Swift Ultra-Violet/Optical Telescope}, {\it Space Science
  Reviews\/}, {\bf 120}(3-4), 95--142.

\bibitem[Rowlinson et~al.(2014)Rowlinson, Gompertz, Dainotti, O'Brien, Wijers,
  \& van~der Horst]{10.1093/mnras/stu1277}
Rowlinson, A., Gompertz, B.~P., Dainotti, M., O'Brien, P.~T., Wijers, R. A.
  M.~J., \& van~der Horst, A.~J., 2014.
\newblock {Constraining properties of GRB magnetar central engines using the
  observed plateau luminosity and duration correlation}, {\it Monthly Notices
  of the Royal Astronomical Society\/}, {\bf 443}(2), 1779--1787.

\bibitem[Salinas et~al.(2023)Salinas, Pichara, Brahm, Pérez-Galarce, \&
  Mery]{10.1093/mnras/stad1173}
Salinas, H., Pichara, K., Brahm, R., Pérez-Galarce, F., \& Mery, D., 2023.
\newblock {Distinguishing a planetary transit from false positives: a
  Transformer-based classification for planetary transit signals}, {\it Monthly
  Notices of the Royal Astronomical Society\/}, {\bf 522}(3), 3201--3216.

\bibitem[{Sari} et~al.(1999){Sari}, {Piran}, \& {Halpern}]{1999ApJ...519L..17S}
{Sari}, R., {Piran}, T., \& {Halpern}, J.~P., 1999.
\newblock {Jets in Gamma-Ray Bursts}, {\it The Astrophysical Journal\/}, {\bf
  519}(1), L17--L20.

\bibitem[{Schady} et~al.(2009){Schady}, {Baumgartner}, {Beardmore}, {Campana},
  {Curran}, {Guidorzi}, {Kennea}, {Mao}, {Margutti}, {Osborne}, {Page},
  {Romano}, {Siegel}, {Stratta}, \& {Ukwatta}]{2009GCN..9512....1S}
{Schady}, P., {Baumgartner}, W.~H., {Beardmore}, A.~P., {Campana}, S.,
  {Curran}, P.~A., {Guidorzi}, C., {Kennea}, J.~A., {Mao}, J., {Margutti}, R.,
  {Osborne}, J.~P., {Page}, K.~L., {Romano}, P., {Siegel}, M.~H., {Stratta},
  G., \& {Ukwatta}, T.~N., 2009.
\newblock {GRB 090618: Swift detection of a bright burst with optical
  afterglow.}, {\it GRB Coordinates Network\/}, {\bf 9512}, 1.

\bibitem[{Sen} et~al.(2022){Sen}, {Agarwal}, {Chakraborty}, \&
  {Singh}]{2022ExA....53....1S}
{Sen}, S., {Agarwal}, S., {Chakraborty}, P., \& {Singh}, K.~P., 2022.
\newblock {Astronomical big data processing using machine learning: A
  comprehensive review}, {\it Experimental Astronomy\/}, {\bf 53}(1), 1--43.

\bibitem[{Sonbas} et~al.(2016){Sonbas}, {Kennea}, {Marshall}, {Page}, \&
  {Palmer}]{2016GCN.18875....1S}
{Sonbas}, E., {Kennea}, J.~A., {Marshall}, F.~E., {Page}, K.~L., \& {Palmer},
  D.~M., 2016.
\newblock {GRB 160117B: Swift detection of a burst with optical counterpart.},
  {\it GRB Coordinates Network\/}, {\bf 18875}, 1.

\bibitem[{Sutskever} et~al.(2014){Sutskever}, {Vinyals}, \&
  {Le}]{2014arXiv1409.3215S}
{Sutskever}, I., {Vinyals}, O., \& {Le}, Q.~V., 2014.
\newblock {Sequence to Sequence Learning with Neural Networks}, {\it arXiv
  e-prints\/}.

\bibitem[{Tohuvavohu} et~al.(2018){Tohuvavohu}, {Gropp}, {Kennea}, {Krimm},
  {Marshall}, {Page}, {Palmer}, {Sbarufatti}, {Siegel}, \&
  {Simpson}]{2018GCN.23158....1T}
{Tohuvavohu}, A., {Gropp}, J.~D., {Kennea}, J.~A., {Krimm}, H.~A., {Marshall},
  F.~E., {Page}, K.~L., {Palmer}, D.~M., {Sbarufatti}, B., {Siegel}, M.~H., \&
  {Simpson}, K.~K., 2018.
\newblock {GRB 180821A: Swift detection of a burst.}, {\it GRB Coordinates
  Network\/}, {\bf 23158}, 1.

\bibitem[{Usov}(1992)]{1992Natur.357..472U}
{Usov}, V.~V., 1992.
\newblock {Millisecond pulsars with extremely strong magnetic fields as a
  cosmological source of {\ensuremath{\gamma}}-ray bursts}, {\it \nat\/}, {\bf
  357}(6378), 472--474.

\bibitem[van Eerten(2014)]{10.1093/mnras/stu1921}
van Eerten, H.~J., 2014.
\newblock {Gamma-ray burst afterglow plateau break time–luminosity
  correlations favour thick shell models over thin shell models}, {\it Monthly
  Notices of the Royal Astronomical Society\/}, {\bf 445}(3), 2414--2423.

\bibitem[Vaswani et~al.(2017)Vaswani, Shazeer, Parmar, Uszkoreit, Jones, Gomez,
  Kaiser, \& Polosukhin]{NIPS2017_3f5ee243}
Vaswani, A., Shazeer, N., Parmar, N., Uszkoreit, J., Jones, L., Gomez, A.~N.,
  Kaiser, L.~u., \& Polosukhin, I., 2017.
\newblock Attention is all you need, in {\em Advances in Neural Information
  Processing Systems\/}, vol.~30, Curran Associates, Inc.

\bibitem[{Walsh} et~al.(2020){Walsh}, {McBreen}, {Martin-Carrillo}, {Dauser},
  {Wijers}, {Wilms}, {Schaye}, \& {Barret}]{2020A&A...642A..24W}
{Walsh}, S., {McBreen}, S., {Martin-Carrillo}, A., {Dauser}, T., {Wijers}, N.,
  {Wilms}, J., {Schaye}, J., \& {Barret}, D., 2020.
\newblock {Detection capabilities of the Athena X-IFU for the warm-hot
  intergalactic medium using gamma-ray burst X-ray afterglows}, {\it Astronomy
  \& Astrophysics\/}, {\bf 642}, A24.

\bibitem[{Wen} et~al.(2022){Wen}, {Zhou}, {Zhang}, {Chen}, {Ma}, {Yan}, \&
  {Sun}]{2022arXiv220207125W}
{Wen}, Q., {Zhou}, T., {Zhang}, C., {Chen}, W., {Ma}, Z., {Yan}, J., \& {Sun},
  L., 2022.
\newblock {Transformers in Time Series: A Survey}, {\it arXiv e-prints\/}.

\bibitem[{Willingale} et~al.(2007){Willingale}, {O'Brien}, {Osborne}, {Godet},
  {Page}, {Goad}, {Burrows}, {Zhang}, {Rol}, {Gehrels}, \&
  {Chincarini}]{2007ApJ...662.1093W}
{Willingale}, R., {O'Brien}, P.~T., {Osborne}, J.~P., {Godet}, O., {Page},
  K.~L., {Goad}, M.~R., {Burrows}, D.~N., {Zhang}, B., {Rol}, E., {Gehrels},
  N., \& {Chincarini}, G., 2007.
\newblock {Testing the Standard Fireball Model of Gamma-Ray Bursts Using Late
  X-Ray Afterglows Measured by Swift}, {\it The Astrophysical Journal\/}, {\bf
  662}(2), 1093--1110.

\bibitem[{Woosley}(1993)]{1993ApJ...405..273W}
{Woosley}, S.~E., 1993.
\newblock {Gamma-Ray Bursts from Stellar Mass Accretion Disks around Black
  Holes}, {\it The Astrophysical Journal\/}, {\bf 405}, 273.

\bibitem[{Wu} et~al.(2019){Wu}, {Liu}, {Bae}, {Chow}, {Iyengar}, {Pu}, {Wei},
  {Yu}, \& {Zhang}]{2019arXiv190806477W}
{Wu}, Y., {Liu}, L., {Bae}, J., {Chow}, K.-H., {Iyengar}, A., {Pu}, C., {Wei},
  W., {Yu}, L., \& {Zhang}, Q., 2019.
\newblock {Demystifying Learning Rate Policies for High Accuracy Training of
  Deep Neural Networks}, {\it arXiv e-prints\/}.

\bibitem[Xiong et~al.(2020)Xiong, Yang, He, Zheng, Zheng, Xing, Zhang, Lan,
  Wang, \& Liu]{xiong2020layer}
Xiong, R., Yang, Y., He, D., Zheng, K., Zheng, S., Xing, C., Zhang, H., Lan,
  Y., Wang, L., \& Liu, T., 2020.
\newblock On layer normalization in the transformer architecture, in {\em
  International Conference on Machine Learning\/}, pp. 10524--10533, PMLR.

\bibitem[Zhang \& Meszaros(2004)]{zhang2004gamma}
Zhang, B. \& Meszaros, P., 2004.
\newblock Gamma-ray bursts: progress, problems \& prospects, {\it International
  Journal of Modern Physics A\/}, {\bf 19}(15), 2385--2472.

\bibitem[Zhang \& Mészáros(2001)]{Zhang_2001}
Zhang, B. \& Mészáros, P., 2001.
\newblock Gamma-ray burst afterglow with continuous energy injection: Signature
  of a highly magnetized millisecond pulsar, {\it The Astrophysical Journal\/},
  {\bf 552}(1), L35.

\end{thebibliography}

\bsp	%
\label{lastpage}
\end{document}